\documentclass[a4paper,twoside,english,aps,preprint,showpacs]{revtex4}
\usepackage[T1]{fontenc}
\usepackage[latin1]{inputenc}
\usepackage{subfigure}
\usepackage{amsmath}
\usepackage{graphicx}
\usepackage{amssymb}

\makeatletter

\newcommand{\noun}[1]{\textsc{#1}}
\providecommand{\tabularnewline}{\\}

\usepackage{babel}
\makeatother
\begin{document}

\title{Ultracold Few-Boson Systems in a Double-Well Trap}

\author{Sascha Zöllner}

\email{sascha@pci.uni-heidelberg.de}

\affiliation{Theoretische Chemie, Institut f\"{u}r Physikalische Chemie, Universit\"{a}t
Heidelberg, INF 229, 69120 Heidelberg, Germany }

\author{Hans-Dieter Meyer}

\email{dieter@pci.uni-heidelberg.de}

\affiliation{Theoretische Chemie, Institut f\"{u}r Physikalische Chemie, Universit\"{a}t
Heidelberg, INF 229, 69120 Heidelberg, Germany}

\author{Peter Schmelcher}

\email{peter@pci.uni-heidelberg.de}

\affiliation{Theoretische Chemie, Institut f\"{u}r Physikalische Chemie, Universit\"{a}t
Heidelberg, INF 229, 69120 Heidelberg, Germany}

\affiliation{Physikalisches Institut, Universit\"{a}t Heidelberg, Philosophenweg
12, 69120 Heidelberg, Germany}

\date{5/24/2006}

\pacs{03.75.Hh, 03.65.Ge, 03.75.Nt}

\begin{abstract}
We investigate the transition of a quasi-one-dimensional few-boson
system from a weakly correlated to a fragmented and finally a fermionized
ground state. Our numerically exact analysis, based on a multi-configurational
method, explores the interplay between different shapes of external
and inter-particle forces. Specifically, we demonstrate that the addition
of a central barrier to an otherwise harmonic trap may supports the
system's fragmentation, with a symmetry-induced distinction between
even and odd atom numbers. Moreover, the impact of inhomogeneous interactions
is studied, where the effective coupling strength is spatially modulated.
It is laid out how the ground state can be displaced in a controlled
way depending on the trap and the degree of modulation. We present
the one- and two-body densities and, beyond that, highlight the role
of correlations on the basis of the natural occupations. 
\end{abstract}
\maketitle

\section{Introduction}

Ever since the first realization of a Bose-Einstein condensate, trapped
ultracold atoms have been the focal point of an enormous number of
research efforts, both from the experimental and the theoretical side
\cite{pitaevskii,dalfovo99,pethick,leggett01}. Allowing for an undreamt-of
level of control, regarding the adjustment of the external as well
as the inter-particle forces via electromagnetic fields, they may
serve as some kind of Rosetta stone for various research areas, ranging
from solid-state physics to optics.

A special focus rests on the aspect of a system's dimensionality.
Particles restricted to a lower-dimensional subspace, such as a wave
guide in the one-dimensional case, often unveil features that are
conspicuously different from isotropic ones. A striking example is
the scattering theory of an ultracold system whose transverse degrees
of freedom are frozen such that an effective one-dimensional description
becomes possible, as developed by Olshanii \cite{Olshanii1998a}.
In that case, the effective interaction strength of the system may
be tuned at will from a weakly correlated to a strongly repulsive
system to an attractive one by merely changing the lengthscale of
the transverse confinement. A particularly absorbing issue is the
so-called Tonks-Girardeau gas of impenetrable bosons \cite{girardeau60},
which provides an analogy to an ideal gas of fermions and has recently
been realized experimentally \cite{kinoshita04,paredes04}.

Traditionally, the physics of ultracold atoms has been studied extensively
on the premise of large numbers of atoms $N$ and sufficiently small
interaction. This legitimates the use of the Gross-Pitaevskii equation
\cite{pitaevskii}, which rests on the assumption of a \emph{macroscopic}
wave function composed of a single orbital. An extension of this idea
to more than one orbital, the so-called \emph{best mean field} \cite{cederbaum03},
has indeed proven to be very efficient in giving a qualitative picture
of the pathway from condensates to fragmentation, which occurs when
the interactions become strong enough to deplete the single-orbital
{}`condensate' \emph{}\cite{alon05}.

Still, there are several rationales to consider systems of \emph{few}
atoms, typically $N\sim1-100$. For one thing, the interesting situation
of a strongly correlated gas is experimentally accessible only for
small systems. Yet a more fundamental argument is that few-particle
systems permit a much higher level of control. There is no thermal
cloud, as for large $N$, associated with decoherence and energetically
dense manifolds of excitations, but a pure quantum system. Instead,
finite-size effects become relevant, and two-body correlations have
to be taken into account from the start. Conversely, few-body systems
are indeed amenable to ab-initio calculations, making it tempting
to analyze their features in detail and without resort to uncontrolled
approximations.

While the last statement is true in principle, it is generally highly
challenging from a computational standpoint. In fact, many attempts
to study few-body systems are based either on analytic solutions for
simple model systems \cite{Busch98,cirone01,hao06,sakmann05,idziaszek06}
or are restricted to very few atoms \cite{Blume2002a,bolda03,tiesinga00,idziaszek:050701}.
Moreover, some numerically exact methods have been put forward recently.
Part of these are actually designed for larger systems but include
only few orbitals as in double-well traps \cite{masiello05,streltsov06},
while others regard generic model systems such as the simple harmonic
oscillator \cite{deuretzbacher06,sundholm04,haugset97} or a double
well \cite{klaiman06}.

The goal of this paper is to investigate the interplay between external
forces, on the one hand, and the effect of manipulating the inter-particle
forces. Our investigation focuses on the numerically exact ground
state obtained via the Multi-Configuration Time-Dependent Hartree
method \cite{mey03:251,mey98:3011,bec00:1}. We consider the example
of $N=2,\dots,6$ one-dimensional repulsive bosons in a double well,
whose barrier separating both wells can be adjusted so that both a
purely harmonic trap as well as large barriers are accessible. We
analyze how this competes with the effect of an increasing interaction
strength, which leads to \emph{fragmentation} and finally \emph{fermionization}
of the ensemble. That interplay is taken one step further by considering
a setup where the interaction is \emph{inhomogeneous}, i.e., the inter-particle
forces depend on the position of a collision, too. This may prove
to be a valuable tool on the road to extracting single atoms from
an ensemble in a controlled way \cite{kuhr01,mohring05}. If the interaction
strength is slightly higher on one side of the trap, the ground state
can be shown to be displaced to the other side. The nature of this
displacement and its dependency on the trap configuration as well
as the interaction's strength and modulation are studied.

This article is organized as follows. In Sec.~\ref{sec:model}, the
model is introduced and the relevant parameter regimes are discussed.
Sec.~\ref{sec:method} contains a concise introduction to the computational
method and how it can be applied to our problem. In the subsequent
section the few-boson system is studied for standard, homogeneous
interactions, casting light on the passage between the low- and strong-correlation
regime, and what effect the trap configuration has. The one- and two-particle
densities are displayed in Secs.~\ref{sub:HO}/\ref{sub:DW}, while
the deeper role of fragmentation is highlighted in \ref{sub:fragmentation}.
The same program is carried out for the case of a collisionally inhomogeneous
setup in Sec.~\ref{sec:gx}.

\section{The model \label{sec:model}}

\newcommand{\exv}[1]{\langle#1\rangle}
 In this article we investigate a system of \emph{few} interacting
bosons ($N=2,\dots,6$) in an external trap. These particles, typically
atoms with mass $M$, are taken to be one-dimensional (1D)---more
precisely, we assume the other two degrees of freedom to be frozen
out in a sense described below.

Let the original 3D system be modeled by the Hamiltonian \[
H=\sum_{i}h_{i}+\sum_{i<j}V(\mathbf{r}_{i}-\mathbf{r}_{j}),\]
where $h=\frac{1}{2M}\mathbf{p}^{2}+U(\mathbf{r})$ is the one-particle
(1P) Hamiltonian with a trapping potential $U$, while $V$ is the
two-particle interaction potential with certain low-energy scattering
parameters $a_{0},r_{0}$ (scattering length and effective range,
respectively). It is well-known that for sufficiently low momenta
($kr_{0}\ll1)$ the details of the interaction become irrelevant.
More precisely, if any other interaction potential is taken, the energy
as well as the asymptotic wave function will remain unchanged as long
as $a_{0}$ agrees (and $r_{0}$, in the next order). In particular,
it is usually convenient to model $V$ by an effective point interaction,
the so-called regularized \emph{pseudo-potential} \cite{huang57}\[
V(\mathbf{r})=\frac{4\pi a_{0}}{M}\delta(\mathbf{r})\partial_{r}r.\]

\subsection{Effective 1D Hamiltonian}

We are now interested in the quasi-1D case, where the trap is supposed
to be anisotropic such that there is a {}`transversal' direction
$(\perp)$ with a characteristic length $a_{\perp}$ much smaller
than that of the longitudinal direction, $a_{\parallel}$. In other
words, the transverse energy gap be sufficiently large compared to
the accessible energy of the system, so that $(\parallel)$ may be
regarded as virtually the only degree of freedom. In this case, one
may integrate out the `frozen' transversal subsystem so as to obtain
an effective 1D interaction \cite{Olshanii1998a}\[
V(x)=g_{\mathrm{1D}}\delta(x),\textrm{ with }g_{\mathrm{1D}}=\frac{4a_{0}}{Ma_{\perp}^{2}}\left(1-\mathcal{C}\frac{a_{0}}{a_{\perp}}\right)^{-1}\quad(\mathcal{C}=1.4603\dots).\]
(These results were derived on the premise of a harmonic transverse
trap potential, that is, $a_{\perp}=1/\sqrt{M\omega_{\perp}}$.) It
is this effective interaction that shall serve as the base for our
investigations, yet with two qualifications. Firstly, the homogeneity
of $V$ will be abandoned later in Sec.~\ref{sec:gx}, where that
modification is discussed in detail. Secondly, though this delta-type
potential is convenient and yields an immediate connection to the
experimentally relevant parameters, it is notoriously intractable
in ab-initio computations like ours. This is a fundamental fact: By
construction, the pseudopotential imposes the condition $\psi'(0^{+})-\psi'(0^{-})=2g_{\mathrm{1D}}\psi(0)$
on the derivative of the relative coordinate---i.e., whenever two
particles meet, the wave function behaves like $e^{-\kappa|x|}$.
This non-smoothness is of course unphysical and solely serves to impose
the correct asymptotics on the wave function. In an exact calculation,
when the problem is approximated by $C^{\infty}$-functions, this
leads to convergence problems. It would be much less artificial to
use a more realistic interaction with a non-zero effective range as
a remedy. We thus opt to mollify the delta function, and instead use
$V(x)=g_{\mathrm{1D}}\delta_{\sigma}(x)$ with the normalized Gaussian\[
\delta_{\sigma}(x)=\frac{\exp\left(-\frac{x^{2}}{2\sigma^{2}}\right)}{\sqrt{2\pi}\sigma},\]
which tends to $\delta$ as $\sigma\to0$ in the distribution sense.
In both analytic and numerical model calculations, we ascertained
that for $\sigma\ll1/|g_{\mathrm{1D}}|$ the results are actually
quite close to the limit $\sigma\to0$. On the other hand, the width
$\sigma$ should not be too small so as to accurately sample the Gaussian.
As a trade-off, we choose a fixed value $\sigma/a_{\parallel}=.05$.
In principle, computations for more than one width could be done in
order to extrapolate to zero; however, we will always keep it fixed.

\subsection{Scaling}

For reasons of universality as well as computational aspects, we will
work with a Hamiltonian rescaled to the lengthscale of the 1D-longitudinal
system, $a_{\parallel}$. More specifically, we carry out a global
coordinate transform $Q':=Q/a_{\parallel}$, with $Q\equiv(x_{1},\dots,x_{N})^{T}$,
which leads to\[
\underbrace{H(Q)/\omega_{\parallel}}_{=:H'(Q')}=\sum_{i}\left(-\frac{1}{2}\partial_{i}^{\prime2}+U'(x'_{i})\right)+\sum_{i<j}V'(x_{i}'-x_{j}').\]
Here $\omega_{\parallel}\equiv1/Ma_{\parallel}^{2}$ defines the energy
scale, and $U'(x'):=U(x=x'a_{\parallel})/\omega_{\parallel}$ etc.
denotes the rescaled potential deprived of any dimensionful parameters.
$H'$ naturally lends itself as a convenient working Hamiltonian,
and we will skip any primes in the following sections. 

As an illustration, the 1D point interaction reduces to \begin{equation}
V'(x')=g_{\mathrm{1D}}^{\prime}\delta(x'),\quad g_{\mathrm{1D}}^{\prime}:=\frac{4a'_{0}}{a_{\perp}^{\prime2}}\left(1-\mathcal{C}\frac{a'_{0}}{a'_{\perp}}\right)^{-1}.\label{eq:V1D}\end{equation}
The only relevant parameter of the interaction is thus the scaled
interaction strength, which in turn requires only the knowledge of
the (scaled) scattering length $a_{0}'=a_{0}/a_{\parallel}$ and the
transverse dimension $a_{\perp}'=a_{\perp}/a_{\parallel}$.

\subsection{Parameter regimes}

As mentioned above, two parameters enter our Hamiltonian: $a_{0}'=a_{0}/a_{\parallel}$
and $a_{\perp}'=a_{\perp}/a_{\parallel}$. Both of course depend on 

\begin{itemize}
\item the 1D length scale $a_{\parallel}=1/\sqrt{M\omega_{\parallel}}$
(due to scaling)
\item the scattering length $a_{0}<a_{\parallel}$ of the atomic species
considered (about 100 a.u. for alkalis; only positive values are considered
here).
\item the transversal lengthscale $a_{\perp}\ll a_{\parallel}$. Of course
$a_{\perp}>a_{0}$ is required unless the validity of the {}`bare'
pseudopotential is put into question. We put $a_{\perp}=0.1a_{\parallel}$
for simplicity.
\end{itemize}
According to (\ref{eq:V1D}), $g_{1D}$ does not depend linearly on
$a_{0}$, but rather tends to $+\infty$ as $a_{0}\to a_{\perp}/\mathcal{C}$
from below. In other words, the system becomes strongly correlated
when the scattering length approaches the transverse-confinement scale,
no matter if the 3D system was strongly interacting to begin with.
We restrict our attention to $g\equiv g_{\mathrm{1D}}>0$. Table~\ref{cap:scales}
illustrates the range of values of $a'_{0}$ for different (longitudinal)
trap frequencies $\omega_{\parallel}$, and what $g_{\mathrm{1D}}^{\prime}$
they correspond to for Na/Rb (at fixed $a'_{\perp}=.1$). 

\begin{center}%
\begin{table}
\begin{tabular}{|c|c|c||c|c|}
\hline 
$\omega_{\parallel}/2\pi\mathrm{Hz}$&
$a'_{0}(\mathrm{Na})$&
$g'_{\mathrm{1D}}$&
$a'_{0}(\mathrm{Rb})$&
$g'_{\mathrm{1D}}$\tabularnewline
\hline
\hline 
$10^{2}$&
$1.9\cdot10^{-3}$&
$.78$&
$5\cdot10^{-3}$&
$2.2$\tabularnewline
\hline 
$10^{3}$&
$6\cdot10^{-3}$&
$2.6$&
$1.6\cdot10^{-2}$&
$8.3$\tabularnewline
\hline 
$10^{4}$&
$1.9\cdot10^{-2}$&
$105$&
$5\cdot10^{-2}$&
$77$\tabularnewline
\hline 
$10^{5}$&
$6\cdot10^{-2}$&
$189$&
\emph{$1.6\cdot10^{-1}$}&
\emph{$-48$}\tabularnewline
\hline
\end{tabular}

\caption{Values of the scaled coupling strength $g'_{\mathrm{1D}}$ for Sodium
and Rubidium for different trap frequencies $\omega_{\parallel}/2\pi$
and $a'_{\perp}=.1$. \label{cap:scales}}
\end{table}
\end{center}

\section{Computational method\label{sec:method}}

Our goal is to investigate the ground state of the system introduced
in Sec.~\ref{sec:model} for all relevant interaction strengths in
a numerically \emph{exact} fashion. In other words, our approach is
not to approximate the \emph{problem} by resorting to two-mode or
mean-field descriptions, but rather to approach the \emph{solution}
in a controllable way. It has to be stressed that this is a highly
challenging and time-consuming task, and only few such studies on
ultracold atoms exist even for model systems \cite{streltsov06,deuretzbacher06,masiello05}.
Our approach relies on the Multi-Configuration Time-Dependent Hartree
\noun{(mctdh)} method \cite{mey90:73,bec00:1,mey03:251}, primarily
a wave-packet dynamics code known for its outstanding efficiency in
high-dimensional applications. To be self-contained, a concise introduction
to this tool---and how it can be adapted to our purposes---is presented
in this section.

The underlying idea of MCTDH is to solve the time-dependent Schrödinger
equation\begin{equation}
\left\{ \begin{array}{c}
i\dot{\Psi}=H\Psi\\
\Psi(Q,0)=\Psi_{0}(Q)\end{array}\right.\label{eq:TDSE}\end{equation}
 as an initial-value problem by expansion in terms of direct (or Hartree)
products $\Phi_{J}$:\begin{equation}
\Psi(Q,t)=\sum_{J}A_{J}(t)\Phi_{J}(Q,t)\equiv\sum_{j_{1}=1}^{n_{1}}\ldots\sum_{j_{f}=1}^{n_{f}}A_{j_{1}\ldots j_{f}}(t)\prod_{\kappa=1}^{f}\varphi_{j_{\kappa}}^{(\kappa)}(x_{\kappa},t),\label{eq:mctdh-ansatz}\end{equation}
using a convenient multi-index notation for the configurations, $J=(j_{1}\dots j_{f})$,
where $f$ denotes the number of degrees of freedom and $Q\equiv(x_{1},\dots,x_{f})^{T}$.
The (unknown) \emph{single-particle functions} $\varphi_{j_{\kappa}}^{(\kappa)}$
are in turn represented in a fixed, primitive basis implemented on
a grid. For indistinguishable particles as in our case, the single-particle
functions for each degree $\kappa=1,\dots,N$ are of course identical
in both type and number ($\varphi_{j_{\kappa}}$, with $j_{\kappa}\le n$).

Note that in the above expansion, not only the coefficients $A_{J}$
are time-dependent, but so are the Hartree products $\Phi_{J}$. Using
the Dirac-Frenkel variational principle, one can derive equations
of motion for both $A_{J},\Phi_{J}$ \cite{bec00:1}. Integrating
this differential-equation system allows one to obtain the time evolution
of the system via (\ref{eq:mctdh-ansatz}). Let us emphasize that
the conceptual complication above offers an enormous advantage: the
basis $\{\Phi_{J}(\cdot,t)\}$ is variationally optimal at each time
$t$. Thus it can be kept fairly small, rendering the procedure very
efficient.

It goes without saying that the basis set above is not inherently
permutation symmetric, as would be an obvious demand when dealing
with bosons. However, the symmetry can be enforced by symmetrizing
the coefficients $A_{J}$, even though this turns out to be unnecessary
as far as the ground state is concerned, which is automatically bosonic
\cite{Penrose56}.

The Heidelberg \noun{mctdh} package \cite{mctdh:package}, which
we use, incorporates a significant extension to the basic concept
outlined so far, the so-called \emph{relaxation method} \cite{kos86:223}.
\noun{mctdh} provides a way to not only \emph{propagate} a wave
packet, but also to obtain the lowest \emph{eigenstates} of the system.
The underlying idea is to propagate some wave function $\Psi_{0}$
by the non-unitary $e^{-H\tau}$ (\emph{propagation in imaginary time}.)
As $\tau\to\infty$, this automatically damps out any contribution
but that stemming from the true ground state $|\mathbf{0}\rangle$,\[
e^{-H\tau}\Psi_{0}=\sum_{J}e^{-E_{J}\tau}|J\rangle\langle J|\Psi_{0}\rangle.\]
 In practice, one relies on a more sophisticated scheme termed \emph{improved
relaxation}. Here $\langle\Psi|H-E|\Psi\rangle$ is minimized with
respect to both the coefficients $A_{J}$ and the configurations $\Phi_{J}$.
The equations of motion thus obtained are then solved iteratively
by first solving for $A_{J}(t)$ (by diagonalization of $(\langle\Phi_{J}|H|\Phi_{K}\rangle)$
with fixed $\Phi_{J}$) and then propagating $\Phi_{J}$ in imaginary
time over a short period. The cycle will then be repeated.

As it stands, the effort of this method scales exponentially with
the number of degrees of freedom, $n^{N}$. Just as an illustration,
using $15$ orbitals and $N=5$ requires $7.6\cdot10^{5}$ configurations
$J$. This restricts our analysis in the current setup to about $N=O(10)$,
depending on how decisive correlation effects are. If these are indeed
essential, then it will turn out later that at least $n=N$ orbitals
are needed for qualitative convergence alone, while the true behavior
may necessitate about $15$. By contrast, the dependence on the primitive
basis, and thus on the grid points, is not as severe. In our case,
the grid spacing should of course be small enough to sample the interaction
potential, and we consider a basis set of $75$ harmonic-oscillator
functions.

\section{Bosons in a double well \label{sec:doublewell}}

\begin{figure}
\begin{center}\includegraphics[%
  width=7.5cm,
  keepaspectratio]{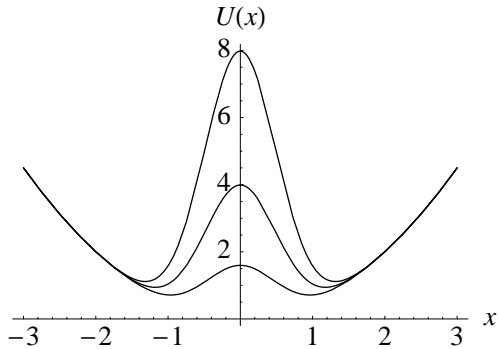}\end{center}

\caption{Sketch of the model potential $U(x)=\frac{1}{2}x^{2}+h\delta_{w}(x)$,
consisting of a harmonic trap plus a normalized Gaussian of width
$w=0.5$ and barrier strengths $h=2,5,10$. \label{cap:DWplot}}
\end{figure}

In this as well as in the following section, we consider the ground-state
properties of bosons in a double-well trap modeled by\[
U(x)=\frac{1}{2}x^{2}+h\delta_{w}(x).\]
This potential is a superposition of a \emph{}harmonic oscillator
(HO), which it equals asymptotically, and a central barrier which
splits the trap into two fragments (Fig.~\ref{cap:DWplot}). The
barrier is shaped as a normalized Gaussian $\delta_{w}$ of width
$w$ and {}`barrier strength' $h$. 

As $w\to0$, the effect of the barrier reduces to that of a mere boundary
condition (since $\delta_{w}\to\delta$), and the corresponding \emph{one-particle}
problem can be solved analytically \cite{Busch03,cirone01}. Although
this soluble borderline case presents a neat calibration, the exact
width $w$ does not play a decisive role, as long as it is larger
than the grid spacing and $w<1$ so as to confine the barrier's effect
to the central region. We choose $w=0.5$ as a trade-off. 

For $h=0$, the case of interacting bosons in a harmonic trap is reproduced.
In Sec.~\ref{sub:HO}, we witness the transition from a simple, weakly
interacting condensate ($g\to0$) to fragmentation and finally the
\emph{Tonks-Girardeau} limit ($g\rightarrow\infty$). As $h\rightarrow\infty$,
the energy barrier will greatly exceed the energy available to the
atoms, and we end up with two \textbf{\emph{}}\emph{isolated wells}.
Higher $g$ then affect only the fragmentation \emph{within} each
of these wells. In between, there is an interesting interplay between
the `static' barrier ($h$) and `dynamical barriers' in the form of
inter-particle forces ($g$). We study this intermediate regime on
the examples of $h\in\{2,5\}$ in Sec.~\ref{sub:DW}.

\subsection{The reference case: $h=0$\label{sub:HO}}

In the absence of a central barrier, we simply deal with a harmonic
trap, which in the case of $N=2$ and point interactions ($\sigma\to0$)
constitutes an exactly solvable problem \cite{Busch98}. Starting
with all interactions turned off ($g=0$), the ground-state solution
is simply the uncorrelated product of the HO-ground state, $\Psi=\phi_{0}^{\otimes N}$.
As long as $g$ is small enough, this behavior remains qualitatively
unaffected by interactions, which only alter the shape of the \emph{single-particle}
functions via a mean field (the Gross-Pitaevskii regime). Increasing
$g$ has the effect of continually depleting the interaction regions
$\{ x_{i}=x_{j}\}$; the repulsion forces the atoms to isolate each
other, a process referred to as \emph{fragmentation} (cf. Sec.~\ref{sub:fragmentation}
for a rigorous account of this). Driven to extremes as $g\to\infty$,
this fragmentation saturates in a so-called \emph{fermionized} state.
In that case the Bose-Fermi mapping \cite{girardeau60} asserts that
letting $g\rightarrow\infty$ emulates the effects of the Pauli exclusion
principle, and the bosons have accomplished to minimize their density
overlap.

\begin{figure}
\subfigure[]{\includegraphics[%
  width=6cm,
  keepaspectratio]{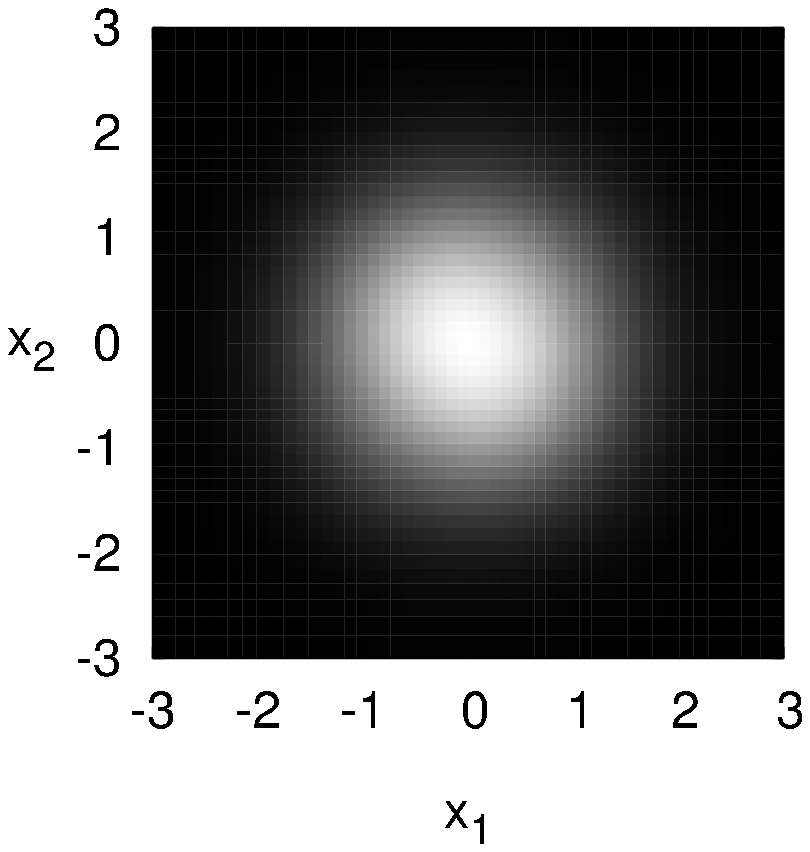}}\subfigure[]{\includegraphics[%
  width=6cm,
  keepaspectratio]{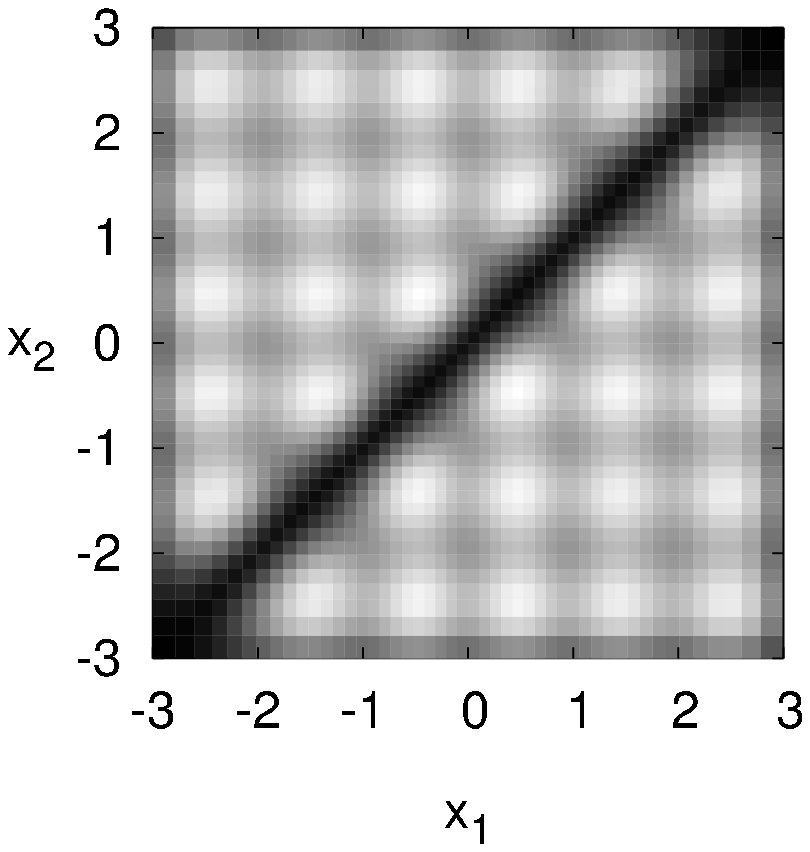}}

\caption{2-particle density $\rho_{2}(x_{1},x_{2})$ for a harmonic trap ($N=6$)
for (a) $g=0.2$ and (b) $g=194$.  \label{cap:density-h0}}
\end{figure}

These qualitative considerations materialize in the reduced densities
of the ground-state. In the 2-particle density $\rho_{2}(x_{1},x_{2})$
---the diagonal kernel of the reduced density operator\[
\rho_{2}:=\mathrm{tr}_{3..N}|\Psi\rangle\langle\Psi|,\]
 which yields the probability density of simultaneously finding any
two particles at $x_{1}$ and $x_{2}$ ---the \emph{correlation diagonal}
$\{ x_{1}=x_{2}\}$ forms an ever deeper dip (the `correlation hole').
This is illustrated in Fig.~\ref{cap:density-h0} for $g=0.2$ and
$g=194$. These represent the borderline cases of a weakly interacting
system and the strong-correlation regime, respectively. In the former
case, the density is simply Gaussian-like. For large enough repulsion,
in turn, the density resembles more and more the checkerboard pattern
produced by a (polarized) fermionic state.%

\begin{center}%
\begin{figure}
\begin{center}\includegraphics[%
  width=8cm,
  keepaspectratio]{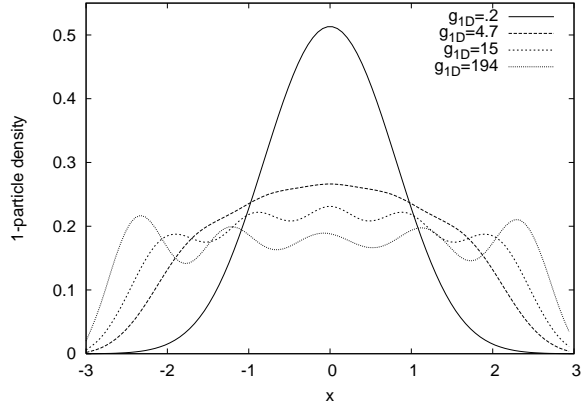}\end{center}

\caption{1-particle density $\rho_{1}(x)$ for a harmonic trap ($N=5$) for
different interactions $g_{\mathrm{1D}}$. Note how the profile changes
from a weakly interacting one ($g=0.2$) to a flattened one due to
fragmentation, and finally to a fermionized profile featuring $N$
humps ($g\ge15$). \label{cap:density1-h0}}
\end{figure}
\end{center}

The 1-particle density $\rho_{1}(x)$ offers another tool to visualize
the fermionization process. Fig.~\ref{cap:density1-h0} gives an
impression of how the profile changes from a harmonic one ($g\ll1$),
to one flattened due to repulsion for mediate $g$, and finally to
the Tonks-Girardeau limit ($g\ge15$), where $N$ humps emerge, mimicking
the fermionic behavior.

\subsection{Central barrier $h>0$\label{sub:DW}}

\begin{center}%
\begin{figure}
\subfigure[]{\includegraphics[%
  width=6cm,
  keepaspectratio]{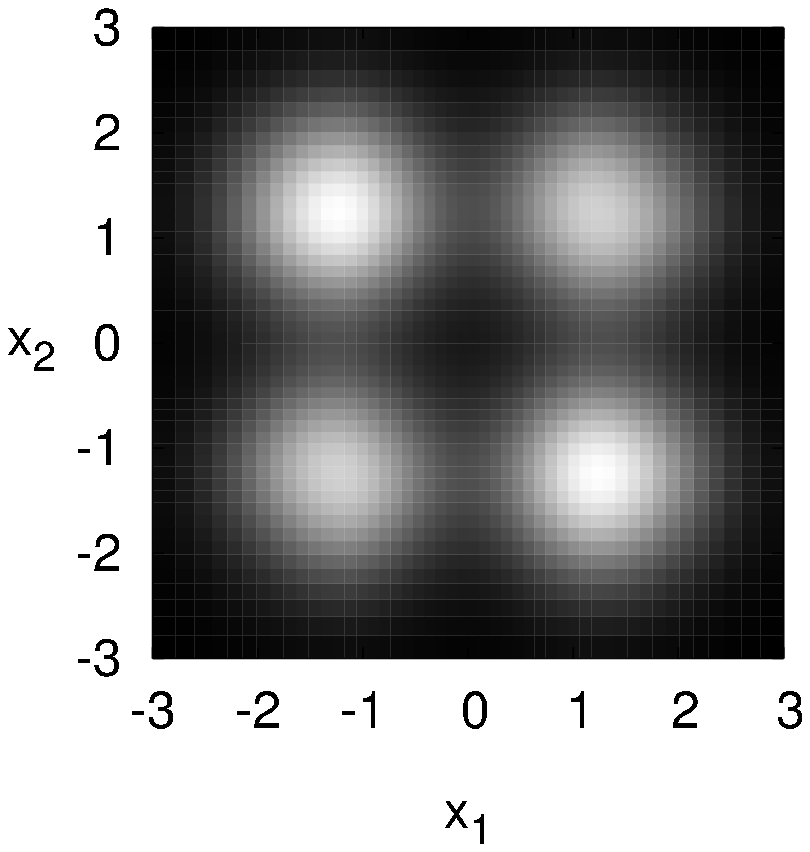}}\subfigure[]{\includegraphics[%
  width=6cm,
  keepaspectratio]{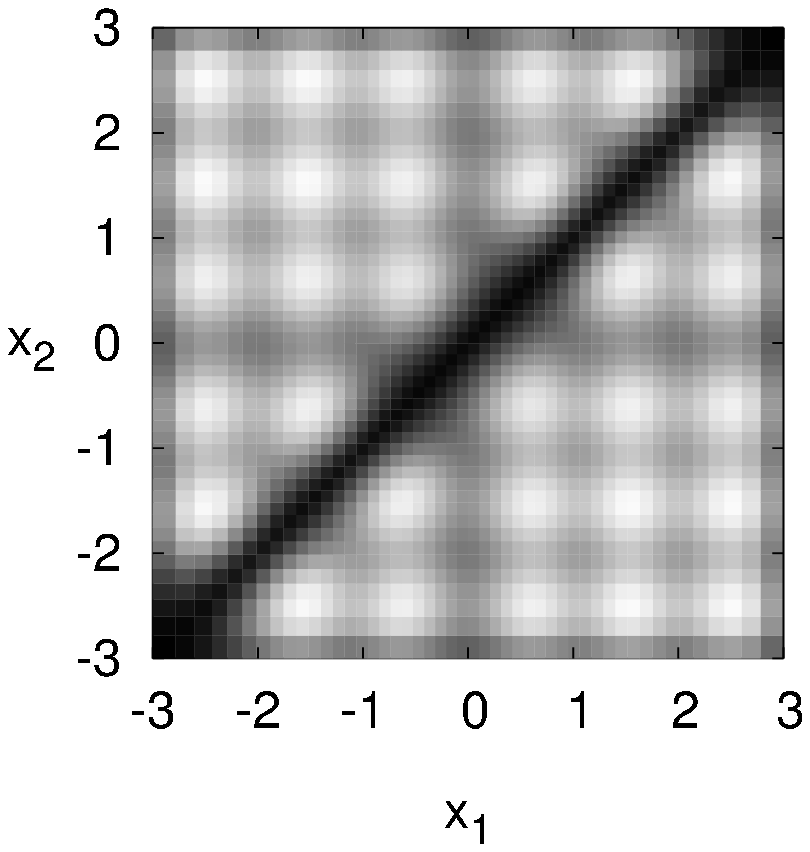}}

\caption{2-particle density $\rho_{2}(x_{1},x_{2})$ for a double-well trap
with barrier strength $h=5$ ($N=6$). (a) The weakly interacting
limit $g=0.2$ (b) The complementary case of fermionization, $g=194$.
\label{cap:density-h5}}
\end{figure}
\end{center}

We now introduce a central barrier $h\delta_{w}(x)$, so as to turn
the harmonic trap into a double well. Upon increasing $g$, there
are now two competing effects, exemplified on the 2-particle density
(Fig.~\ref{cap:density-h5}): For small enough $g$, the barrier
$h$ dispels the atoms from the center $x=0$. This state is \textit{uncorrelated}:
they are localized in both wells regardless of whether there are already
any other ones. Indeed, in Fig.~\ref{cap:density-h5}(a) the diagonal
$\{ x_{1}=x_{2}\}$ shows only a tiny depletion due to repulsion.
In the complementary case $g\rightarrow\infty$, the barrier will
be almost outweighed by the inter-particle repulsion. In this Tonks-Girardeau
limit, the atoms distribute so as to minimize their density overlap,
at the price of an additional potential energy in the barrier region.
Hence the two-particle density in Fig.~\ref{cap:density-h5}(b) differs
from the harmonic counterpart only in the stripes along $x_{1/2}=0$,
which indicate suppression in the central-barrier region. The question
as to what happens in the intermediate region requires a distinction
between even and odd particle numbers; it will be the focus of the
ensuing paragraphs.

\subsubsection{Even $N$: assisted fragmentation}

\begin{figure}
\subfigure[]{\includegraphics[%
  width=8cm,
  keepaspectratio]{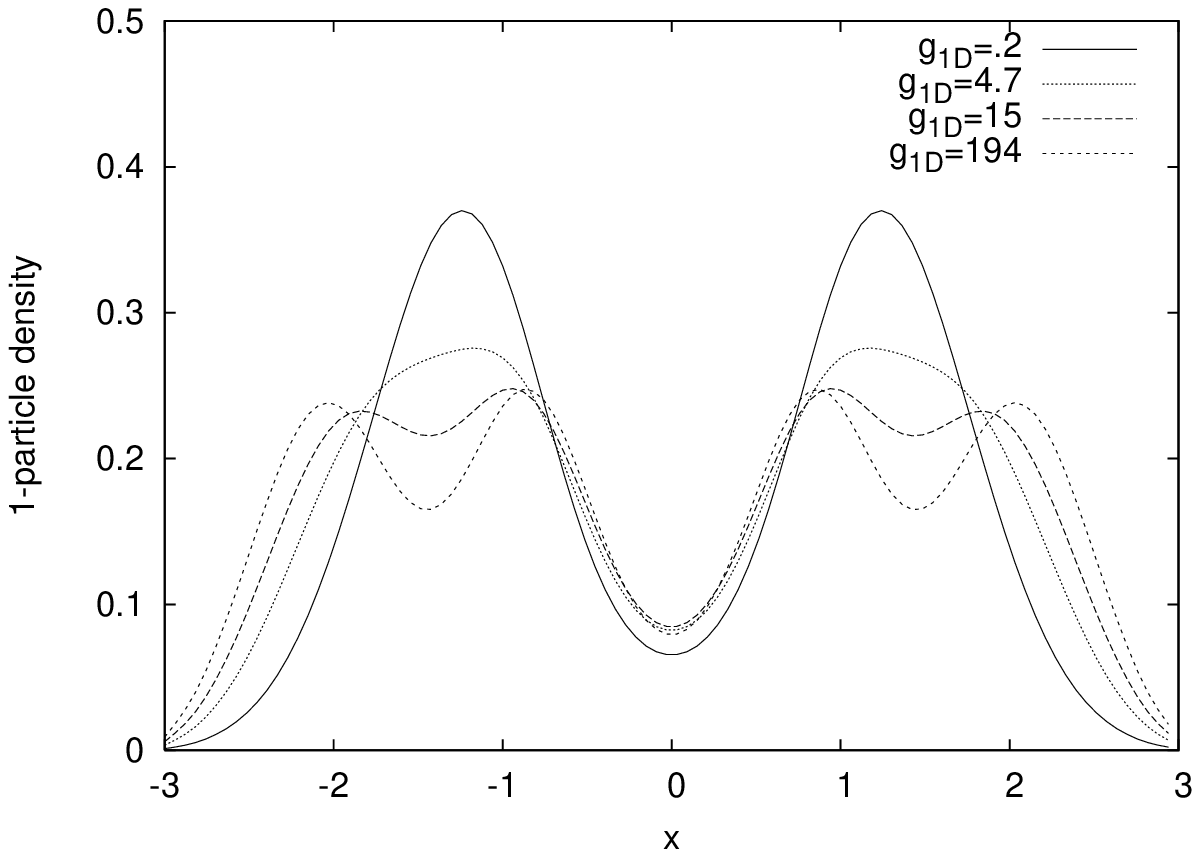}}\subfigure[]{\includegraphics[%
  width=8cm,
  keepaspectratio]{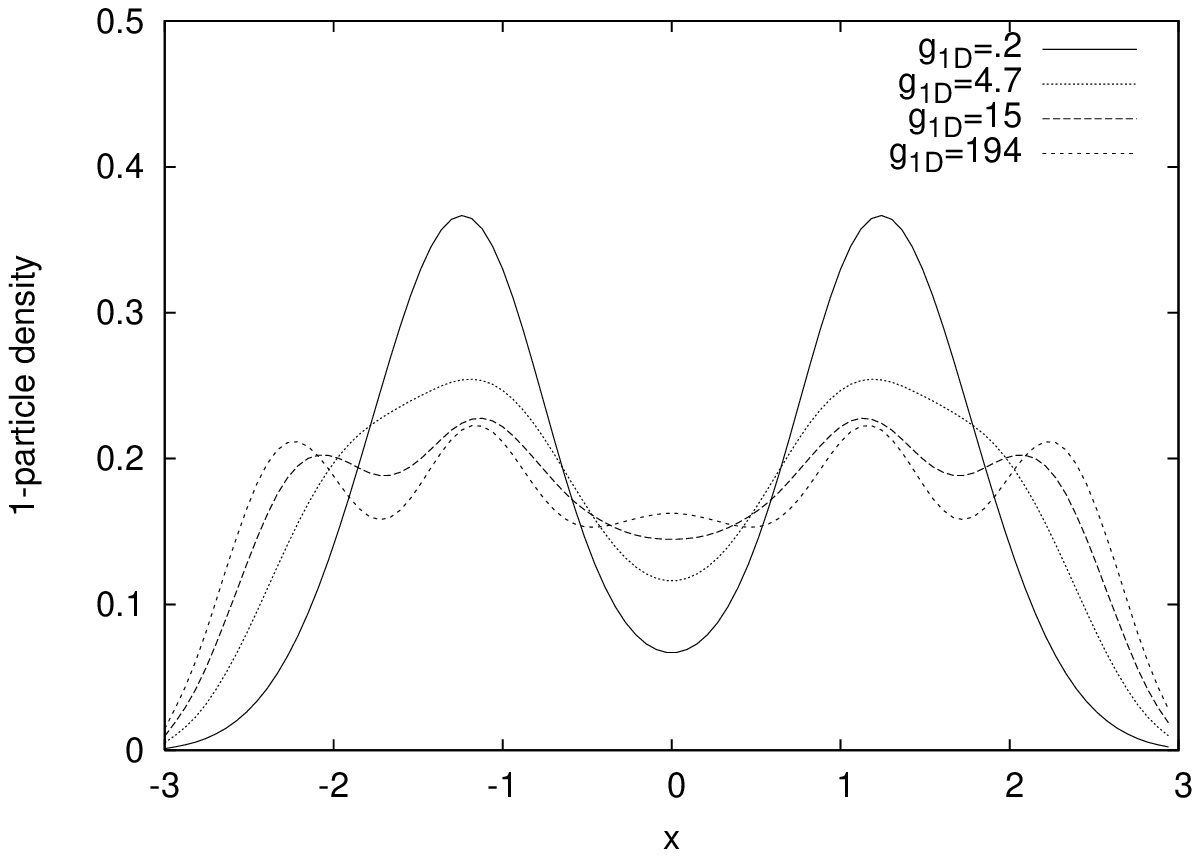}}

\caption{1-particle density in a double well with $h=5$: Even vs. odd number
of atoms. (a) For $N=4$ atoms, the fragmentation process is in part
supported by the central barrier. By contrast, (b) indicates in what
sense fermionization is suppressed for an odd number ($N=5$), since
by symmetry one particle should eventually reside in the center. \label{cap:density1-h5} }
\end{figure}

For an \textit{even} number of atoms---in what follows $N\in\{2,4,6\}$---they
initially (at $g=0$) populate each well with $N/2$ atoms. Upon raising
the interaction strength, they seek to isolate each other, which can
best be done by having a fragmentation \textit{within} each well,
interfering only little with the central barrier. This situation is
depicted in Fig.~\ref{cap:density1-h5} for four atoms. The plots
for $g=4.7,15$ reflect the tipping from a coherent state to fragmentation.
For high enough $g$, the accessible energy approaches that of the
barrier, $Nh/\sqrt{2\pi}w$, so that tunneling becomes more and more
dominant. Beyond that point, the density profile is expected to resemble
that of a purely harmonic trap with slightly suppressed amplitude
in the barrier region. In particular, for any finite $h$, the Tonks-Girardeau
limit will be of a generic form in that it exhibits $N$ density maxima.
In conclusion, one might say that the fragmentation here is assisted
insofar as the central barrier helps isolate the ensemble, a statement
put more precisely in \ref{sub:fragmentation}.

\subsubsection{Odd $N$: delayed fragmentation}

In the case of an \textit{odd} number ($N=3,5$), the situation differs.
At $g=0$, the atoms are again coherently distributed over the two
wells. As we strengthen the interactions, they try to enlarge their
distance---but by symmetry, this can now only be done by placing one
particle at $x=0$, which in turn is impeded by the barrier (cf. Fig.~\ref{cap:density1-h5}b).
In other words, the system will have to pay the added interaction
energy and distribute  the extra particle over the two wells, until
the former one becomes high enough to afford the place in the central-barrier
zone. In that sense, the fragmentation is attenuated by the double-well
trap.

\subsection{Ground-state energy}

\begin{figure}
\begin{center}\includegraphics[%
  width=7cm]{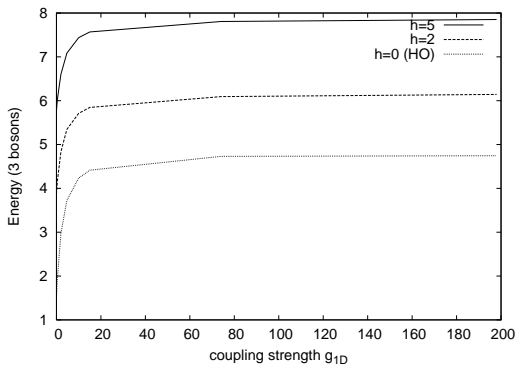}\end{center}

\caption{Energy $E(g)$ for the case $N=3$. Note the slightly different effect
of the interaction, measured by the slope at $g=0$, for different
barrier strengths $h=0$ (harmonic trap) and $h=2,5$. The saturation
as $g\rightarrow\infty$ corresponds about to a fermionized state.\label{cap:energy}}
\end{figure}

One central aspect of our analyses is the ground-state energy. Fig.~\ref{cap:energy}
depicts a typical evolution, $E(g)$, as a function of the coupling
strength. As suggested above, the $g\rightarrow\infty$ state is isomorphic
to a non-interacting fermionic state. In this light, also the ground-state
energy $E(g)$ (Fig.~\ref{cap:energy}) may be interpreted as connecting
the free bosonic ($g=0$) and the free fermionic value, corresponding
to the saturation as $g\rightarrow\infty$. 

The effect of the interaction at $g=0$ can be measured by the slope
\[
\frac{dE}{dg}(0)=\frac{N(N-1)}{2}\langle00|\delta_{\sigma}(x_{1}-x_{2})|00\rangle\stackrel{\sigma\to0}{\sim}\frac{N(N-1)}{2}\int|\phi_{0}(x)|^{4}dx,\]
 given by the density overlap of two atoms in the non-interacting
ground state. The centered harmonic-oscillator orbital $\phi_{\mathrm{HO}}$
by construction has a low curvature (i.e., kinetic energy), thus producing
a rather high density overlap. It is thus more susceptible to the
onset of interactions. By contrast, the presence of a central potential-energy
barrier ($h\to\infty$) evokes an orbital $\phi_{\mathrm{DW}}$ delocalized
in both wells. Its density overlap in turn will be smaller, which
can be seen schematically by assuming for a moment that $\phi_{\mathrm{DW}}(x)\sim\frac{1}{\sqrt{2}}\left[\phi_{\mathrm{HO}}(x-x_{0})+\phi_{\mathrm{HO}}(x+x_{0})\right]$
is built from a HO orbital centered in both minima $\pm x_{0}$. Neglecting
the density overlap between the right- and left-hand contributions,
$\int|\phi_{\mathrm{DW}}|^{4}\simeq\frac{1}{2}\int|\phi_{\mathrm{HO}}|^{4}$,
suggesting that in a double well, the atoms will feel a lesser effect
when interactions are turned on. This can be seen in Fig.~\ref{cap:energy}.

\subsection{The role of fragmentation\label{sub:fragmentation}}

We have so far relied on a more intuitive notion of fragmentation.
It is natural to ask whether some of our previous assertions can be
put more quantitatively. This we seek to do in the present subsection,
not only to highlight some deeper concepts, but also to show why a
numerically exact approach is so vital.

It has been argued that ever stronger interactions introduce correlations
to the system, viz., the one-particle Hamiltonian will be no longer
dominant, but the influence of the two-particle operator $V$ takes
over. The latter one imprints explicit two-body correlation terms
on the eigenvector, as illustrated, for instance, on the example of
two atoms in a trap \cite{Busch98,cirone01} or the fermionization
limit $g\to\infty$ solvable for any $N$ \cite{girardeau01} starting
from the Bose-Fermi mapping,\[
\Psi(Q)\propto e^{-|Q|^{2}/2}\negthickspace\prod_{1\le i<j\le N}\negthickspace|x_{i}-x_{j}|.\]
Technically speaking, it is therefore far from clear how long $\Psi$
can be well approximated by states composed of one-particle functions,
let alone a single such configuration as in mean-field approaches
(see \cite{alon05} and references therein). 

Being exact, our method offers a handle on these convergence questions,
based on a criterion put forward by Penrose \cite{Penrose56}. Consider
the spectral decomposition of the one-particle density matrix \begin{equation}
\rho_{1}=\mathrm{tr}_{2..N}|\Psi\rangle\langle\Psi|\equiv\sum_{a}n_{a}|\phi_{a}\rangle\langle\phi_{a}|,\label{eq:rho1}\end{equation}
 where $n_{a}\in[0,1]$ is said to be the population of the \textit{natural
orbital} $\phi_{a}$. Obviously, if all $n'_{a}\equiv n_{a}N\in\mathbb{N}$
($\sum_{a}n_{a}'=N$), then the density may be mapped to the (non-interacting)
number state $|n'_{0},n'_{1},\dots\rangle$ based on the \emph{natural}
one-particle basis; for non-integer values it extends that concept.
In particular, the highest such occupation, $n_{0}$, may serve as
a measure of \emph{non}-fragmentation: for $n_{0}=1$, a simple condensate
is recovered. This is the well-known borderline case of the Gross-Pitaevskii
eq.: as $g\rightarrow0$, $\rho_{1}\rightarrow|\phi_{0}\rangle\langle\phi_{0}|$
\cite{lieb03}. The complementary fermionization limit ($g\rightarrow\infty$)
has been investigated semi-analytically drawing on the Fermi/Bose
mapping, yielding $n_{0}\simeq N^{-.41}$ for a harmonic trap \cite{girardeau01}. 

\begin{figure}
\begin{center}\subfigure[]{\includegraphics[%
  width=8cm,
  keepaspectratio]{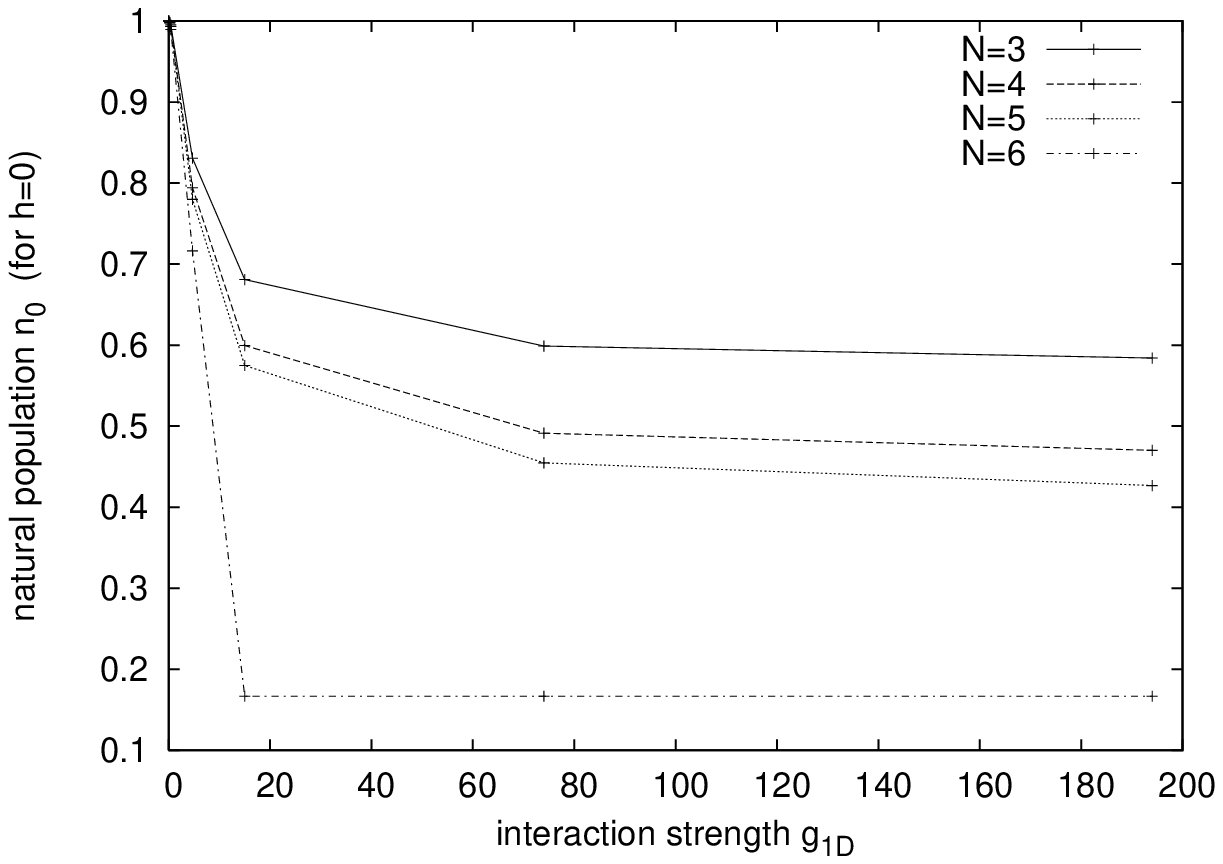}}\subfigure[]{\includegraphics[%
  width=8cm,
  keepaspectratio]{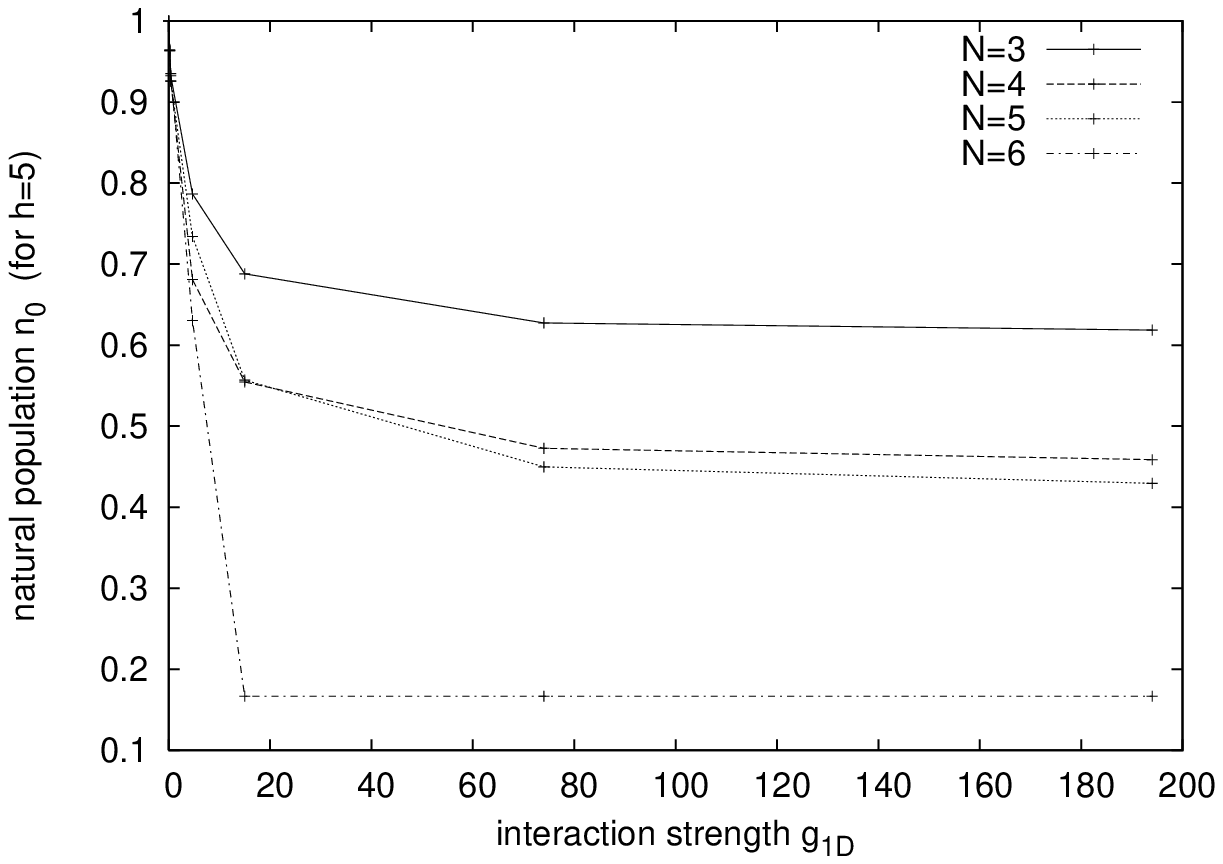}}\end{center}

\caption{Maximum natural population $n_{0}(g)$ for (a) a harmonic trap ($h=0$)
and (b) a double well ($h=5$). For even $N$, the fragmentation is
enhanced by the higher barrier, whereas for odd $N$ it is slowed.
\label{cap:fragm}}
\end{figure}

In between those extremes, a thorough many-body treatment is indispensable.
This is illustrated in Fig.~\ref{cap:fragm}, where $n_{0}(g)$ is
plotted. Observe that, for one thing, the high-correlation value $n_{0}(g\to\infty)$
is much too low for $N=6$ compared with the above value $n_{0}\simeq.48$,
a numerical effect explained below. Moreover, the lines for $N=4$
and $N=5$ in the double-well trap (Fig.~\ref{cap:fragm}b) are partly
reversed as compared to the harmonic trap, while the $N=3$ value
of $n_{0}$ is shifted upward. This reflects the suppression of fragmentation
for odd atom numbers with respect to even $N$.

\begin{figure}
\begin{center}\subfigure[]{\includegraphics[%
  width=6cm]{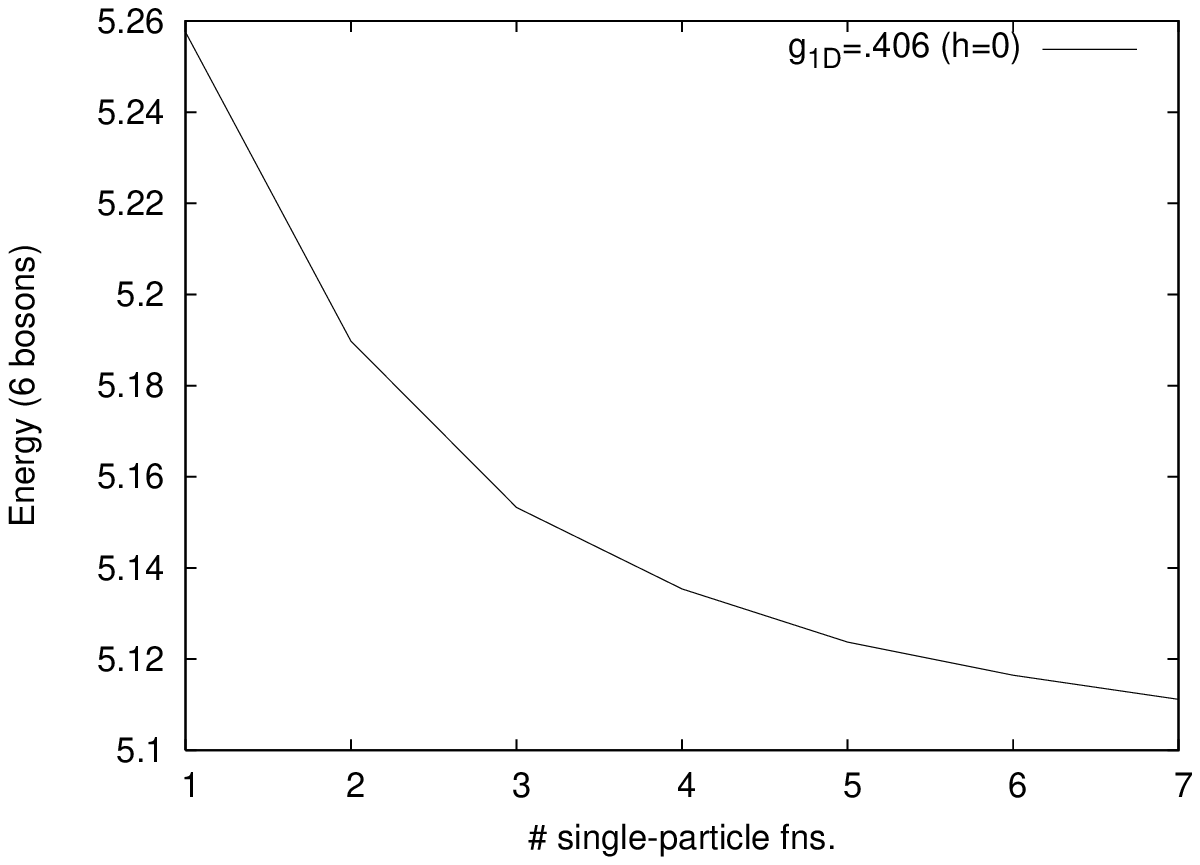}}\subfigure[]{\includegraphics[%
  width=6cm]{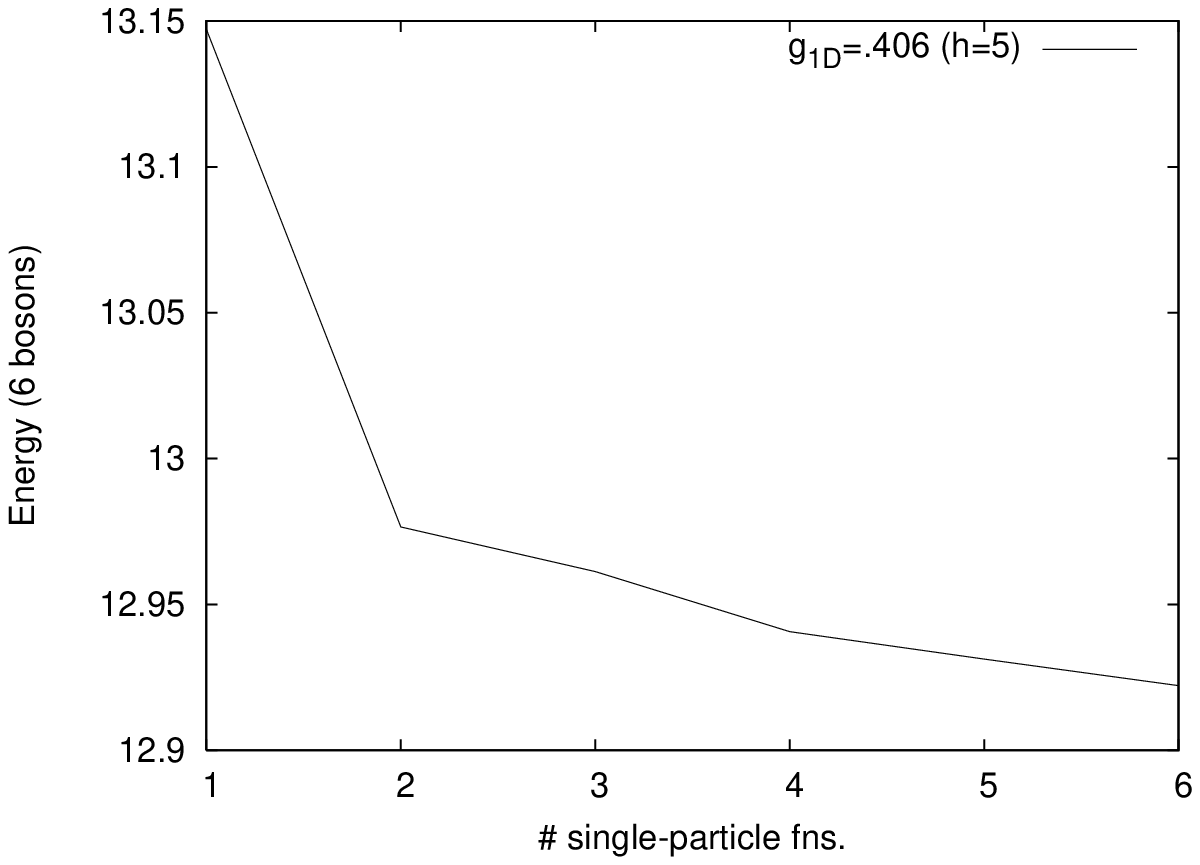}}\subfigure[]{\includegraphics[%
  width=6cm]{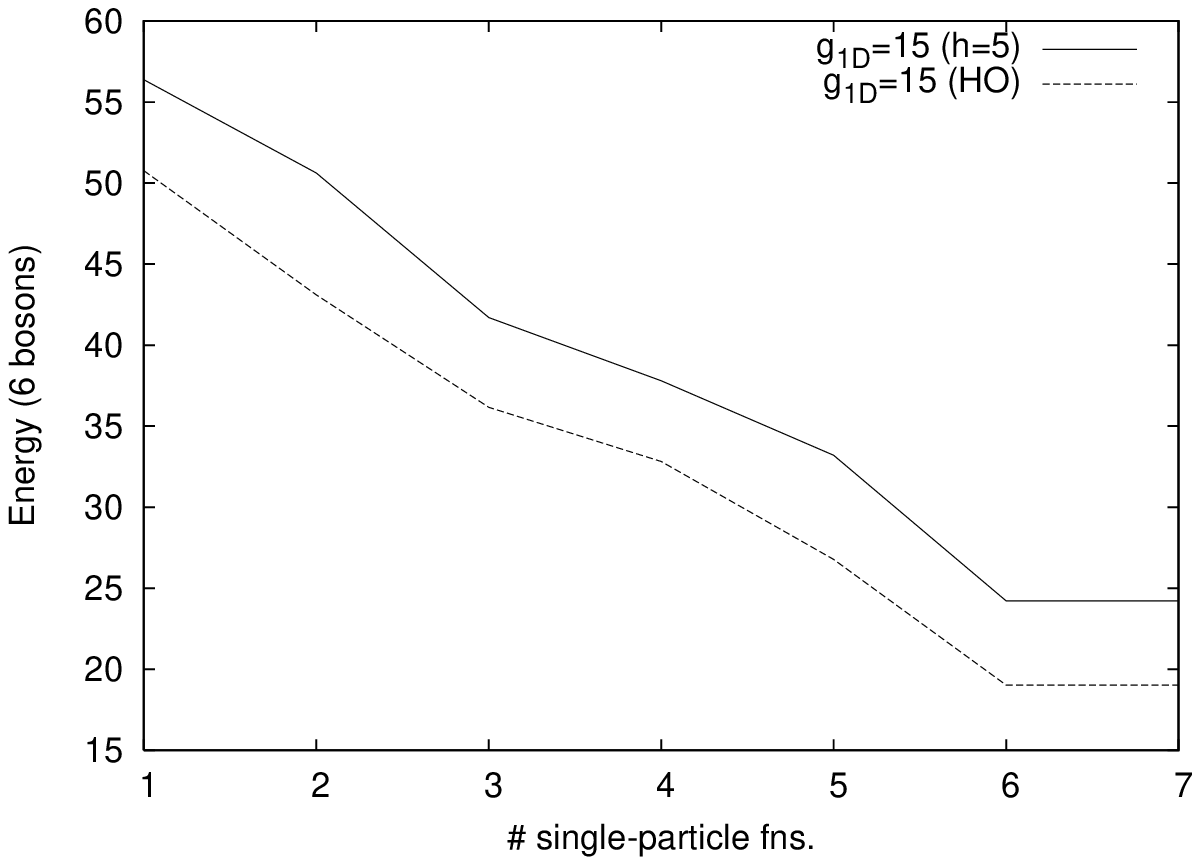}}\end{center}

\caption{Convergence of the energy $E(n)$ in the number of single-particle
functions $n$ for six bosons. For weak interaction $g=.406$, the
difference between the harmonic (a) and the split trap (b) is conspicuous,
though in both cases Gross-Pitaevskii ($n=1$) is not too far off.
As opposed to that, the onset of fermionization ($g=15$) wipes out
that difference up to a residual energy offset due to the central
barrier (c). Here $n=N$ orbitals are needed to be at least qualitatively
correct. \label{cap:En}}
\end{figure}

The role played by correlations may also be highlighted by the number
of single-particle functions needed in a calculation, $n$ (cf. \ref{sec:method}).
It also determines how many terms in the spectral expansion of $\rho_{1}$
(Eq. \ref{eq:rho1}) are included. Figures~\ref{cap:En}(a-c) show
the convergence of the ground-state energy of $N=6$ bosons as $n$
is varied. For rather low $g=.406$, the convergence for the harmonic
trap is fairly smooth, and the variations are altogether relatively
small. This asserts that Gross-Pitaevskii works qualitatively well,
though it may already take many orbitals to achieve a high accuracy.
For a double well with $h=5$, the convergence is much more abrupt.
Adding just another orbital, $n=2$, lowers the energy drastically
in comparison with any further refinements. This is intelligible,
regarding the fact that two-mode models based essentially on the anti-/symmetric
orbitals of a double well are widely used in that context.

Fig.~\ref{cap:En}(c) in turn casts a light on the strong-correlation
regime, $g=15$ (in fact its ground state for $h=5$ is already atop
the central energy barrier.) Here the lines $E(n)$ for $h=0,5$ are
almost parallel, shifted only by the energy offset of about $h\sim\int h\delta_{w}(x)dx$.
Clearly simple mean-field theory is off by more than a factor of 2,
as it drastically overestimates the interaction energy and fails to
reproduce the characteristic Tonks-Girardeau profile with $N$ humps.
Just when $n=N=6$, the convergence suddenly settles, and the \emph{qualitatively}
correct behavior can be observed. Interestingly enough, adding another
orbital, $n=7$, has virtually no effect on that energy scale: the
first $N$ natural occupations all turn out to be $n_{0}=1/N$, corresponding
to a number state $|1_{0},\dots,1_{N-1}\rangle$ in the natural-orbital
basis. This is of course \emph{quantitatively} incorrect in light
of the expected behavior $n_{0}\sim N^{-0.41}$, which really can
be recovered if $n\gtrsim15$ single-particle functions are included.
But apparently that ground state seems to be a somewhat stable \emph{intermediary}
solution in the fermionization limit if the subspace $\mathrm{span}\{\Phi_{J}\mid j_{\kappa}\le n\}$
is taken too narrow. This provides an insightful link to multi-orbital
approaches \cite{alon05,cederbaum03}. %

Let us remark that the criterion employed above is conventional and
amenable, but not imperative. Indeed, the two-particle density $\rho_{2}$
by nature reveals correlation effects even more clearly.

\section{Inhomogeneous interactions \label{sec:gx}}

We have so far relied on the assumption of \emph{homogeneous} two-particle
forces. These are invariant under global translations and thus depend
on $x_{i}-x_{j}$ alone. While this premise is most natural from a
fundamental point of view, we should keep in mind that our description
is not a \emph{fully} microscopic one, even if we ignore the internal
structure of the underlying atoms. Rather, it is an effective model
stripped not only of the transverse degrees of freedom, but of course
also of the electromagnetic fields that manipulate both external and
inter-particle forces. 

With this in mind, it appears legitimate to conceive situations where
the strength of the interactions depends in addition on the position
where the collision takes place, as was done in a mean-field framework
in Ref.~\cite{theocharis05} (see also citations therein). This may
be induced by means of a Feshbach resonance, tuning $a_{0}(\mathbf{B})$
by adding a spatial dependence to the magnetic field. In our one-dimensional
setting, it seems even more convenient to exploit the parametric dependence
on the transverse subsystem, and modify $a_{\perp}$ locally so as
to imprint a spatial dependence on $g_{\mathrm{1D}}$.

Without reference to the specific realization and its concrete experimental
constraints, we perform a case study where a generic model for the
inhomogeneity is assumed to begin with. This model will be presented
in Sec.~\ref{sub:Model-interaction}. The interplay of that dynamical
inhomogeneity with the external forces will be studied for a harmonic
(\ref{sub:mHO}) and a double-well trap (\ref{sub:mDW}).

\subsection{Model interaction\label{sub:Model-interaction}}

Whereas modeling a position-dependent interaction in a mean-field
description (as in \cite{theocharis05}) is straightforward, since
one only has an effective one-particle problem, one faces a conceptual
problem when using a many-body framework. In general, the coupling
would depend on both participants $x_{i},x_{j}$, which is technically
possible if somewhat awkward. For it to make sense intuitively, we
require that its modulation lengthscale be much larger than the {}`radius'
of collision, $\sigma$.

With this is mind, it is natural to model our interaction in terms
of the respective relative coordinate $r:=x_{i}-x_{j}$ (for fixed
$i,j$) and---in order to keep $V$ formally symmetric--- the center
of mass $2R:=x_{i}+x_{j}$: \[
V(r,R)=g(R)\delta_{\sigma}(r).\]
The demand that $g$ change slowly over the support of $V$ can be
cast as\begin{equation}
\left\Vert \frac{g}{g'}\right\Vert _{\infty}\gg\sigma\label{eq:slow}\end{equation}
in the supremum norm. 

\begin{figure}
\begin{center}\includegraphics[%
  width=6cm,
  keepaspectratio]{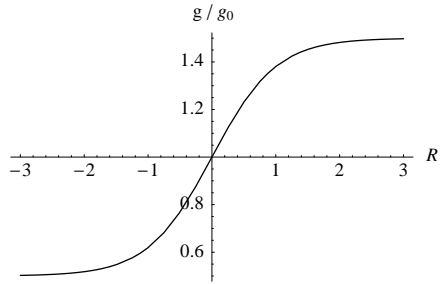}\end{center}

\caption{Our model of the position-dependent coupling $g(R)/g_{0}=1+\alpha\tanh\left(\frac{R}{L}\right)$.
The relative modulation, here $\alpha=0.5$, determines the asymptotic
difference from the average value $g_{0}$, while the modulation length
$L=1$ shall remain fixed. \label{cap:gX}}
\end{figure}

There are various possibilities just what scenario should be examined,
be it some kind of collision-enhanced tunneling or dynamical self-trapping
\cite{theocharis05}. We concentrate on a specific model where $g$
is essentially imbalanced between the right- and left-hand sides of
the trap (Fig.~\ref{cap:gX}):\[
g(R)=g_{0}\left[1+\alpha\tanh\left(\frac{R}{L}\right)\right].\]
This signifies that for $|R|\gg L$, the coupling takes on the asymptotic
values\[
g_{\pm}\equiv\lim_{|R|\to\pm\infty}\negthickspace g(R)=g_{0}(1\pm\alpha),\]
while it changes on a scale of $L$ near the trap's center about $g_{0}$.
The parameter $\alpha$ regulates both the relative difference between
the asymptotic strengths and their ratio: \begin{eqnarray*}
\Delta g & \equiv & |g_{\pm}-g_{0}|=g_{0}\alpha,\\
\frac{g_{+}}{g_{-}} & = & \frac{1+\alpha}{1-\alpha}.\end{eqnarray*}
Eq.~(\ref{eq:slow}) can be met if $L\gg\sigma\alpha$, which is
effortlessly fulfilled if we choose $L=1$ for convenience.%

\subsection{The reference case: $h=0$\label{sub:mHO}}

Generally speaking, the ground state of atoms immersed in a harmonic
trap will be centered at the trap's bottom, assuming that we start
with a weakly interacting ensemble. Hence the modulation of the coupling
strength $g$ beyond the center will pass them largely unnoticed.
It is only for strong enough repulsive interaction---where fragmentation
sets in---that the density profile will start to split and shift partly
outward, thus experiencing an asymmetry. 

\begin{figure}
\begin{center}\includegraphics[%
  width=8cm,
  keepaspectratio]{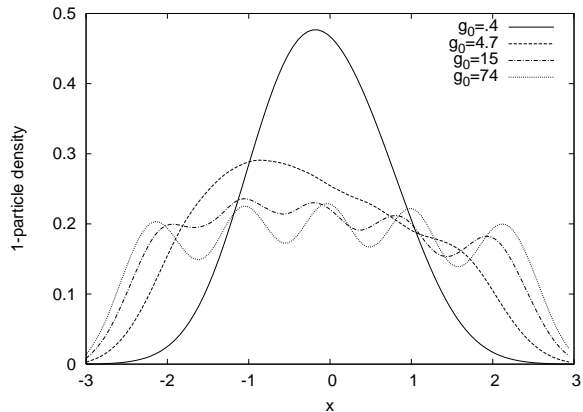}\end{center}

\caption{1-particle densities $\rho_{1}(x)$ for a harmonic trap ($N=5$)
in the case of inhomogeneous interactions, here $\alpha=.5$. The
profile features an imbalance for smaller interactions $g_{0}$, where
the wave packet is centered too much to sample the modulation of $g(R)$.
When fragmentation sets in, the profile splits and the asymmetry becomes
more distinct. In the fermionization limit, the energy costs of an
imbalance become too large to keep it up. \label{cap:density1-mHO}}
\end{figure}

This picture is supported by our calculations, as demonstrated in
Figure~\ref{cap:density1-mHO} for $N=5$ atoms. For low enough $g_{0}=.4$,
measuring the average interaction strength, the harmonic profile is
barely altered from the homogeneous case $\alpha=0$. An imbalance
is noticed for medium $g_{0}=4.7$: the atoms are now able to sample
the modulation of the coupling strength and find it more inexpensive
to locate in the less repulsive zone $x<0$ ($g_{-}$). However, this
effect ceases as the repulsion becomes larger ($g\ge15$). This may
be interpreted as follows: the energetical costs for concentrating
several particles near one spot are soaring, and this \emph{in total}
eventually outweighs the \emph{relative} energy savings reached by
an imbalance.%

\begin{figure}
\subfigure[]{\includegraphics[%
  width=6cm,
  keepaspectratio]{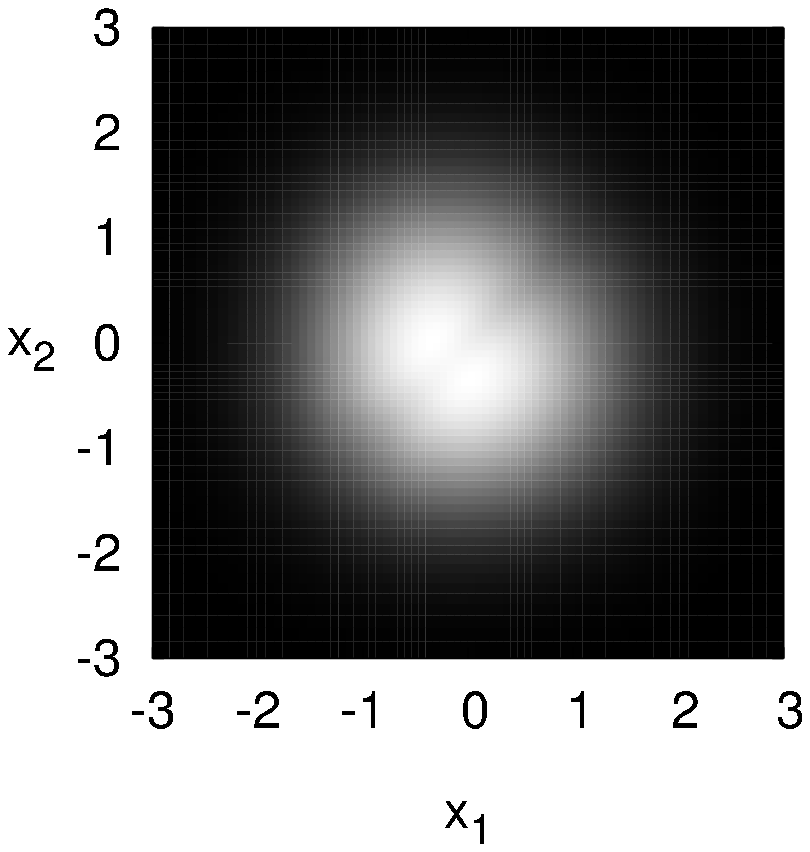}}\subfigure[]{\includegraphics[%
  width=6cm,
  keepaspectratio]{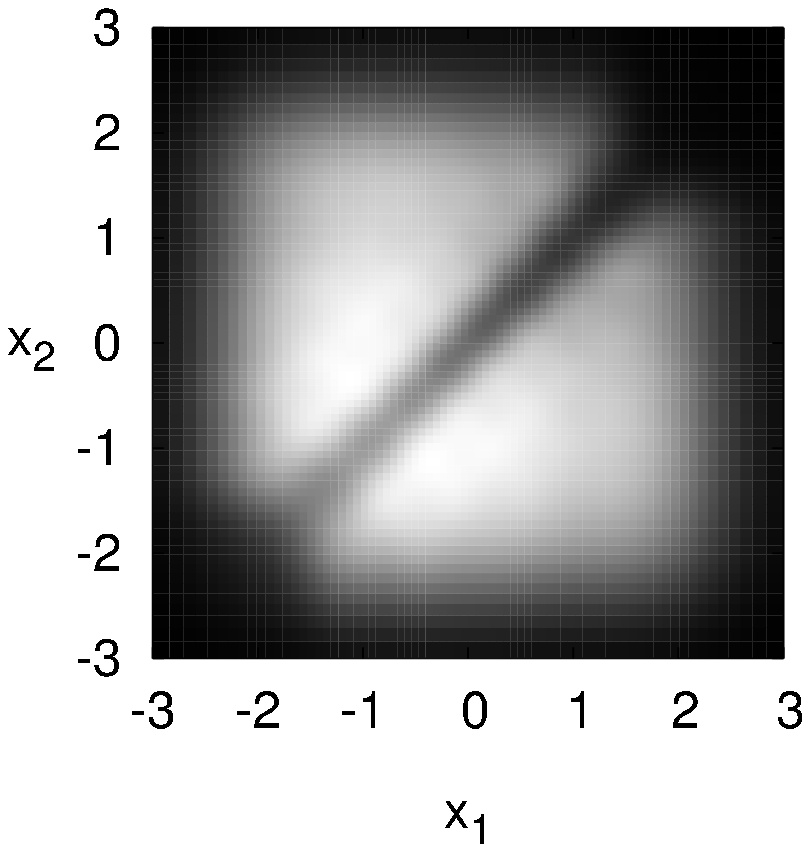}}

\caption{2-particle density for a harmonic trap in the presence of inhomogeneous
interactions ($N=5$). (a) For $g=0.4$, the packet is localized about
the center, thus widely ignoring the modulation. (b) When fragmentation
sets in ($g=15$), it starts to delocalize and consequently shifts
to $R<0$. \label{cap:density-mHO}}
\end{figure}

A look into the two-body correlations $\rho_{2}(x_{1},x_{2})$ in
Fig.~\ref{cap:density-mHO} helps us clarify what happens. Along
the correlation diagonal $\left\{ x_{1}=x_{2}\right\} $, $R=x_{1/2}$
holds. It is here that the modulation can have an impact, whereas
for $x_{1}=-x_{2}$, $g(R)=g(0)=g_{0}$ as usual. Clearly the density
on the diagonal $\{ x_{1}=x_{2}\}$ must be spread enough for the
modulation to become effective. This is not the case for small interactions.
Indeed, for $g=0.4$, the packet is localized about the center, thus
widely ignoring the modulation. Yet for medium $g_{0}$ (Fig.~\ref{cap:density-mHO}b),
the repulsion-driven broadening has become distinct enough so that
the ground states exhibits some left-right asymmetry. For strong fragmentation,
the correlation diagonal will in turn be fully depleted, so obviously
the atoms can no longer realize the modulation, and the very premise
of an inhomogeneity-based displacement has become obsolete.%

\begin{figure}
\begin{center}\includegraphics[%
  width=7cm,
  keepaspectratio]{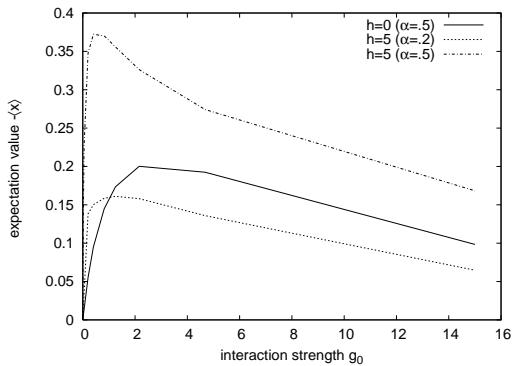}\end{center}

\caption{The ground-state displacement $-\exv{x}$ as a function of the average
interaction $g_{0}$ ($N=5)$. Its universal behavior is an increase
up to a maximum value followed by a slow decay. The increase at $g_{0}=0$
is strongly enhanced in the presence of a barrier $h>0$, while for
the purely harmonic trap ($h=0$), it is rather slow. Of course the
maximum itself is much more pronounced for higher modulations $\alpha$,
while being absent in the homogeneous case $\alpha=0$. \label{cap:mHO-q}}
\end{figure}

The above findings are nicely wrapped up in Fig.~\ref{cap:mHO-q},
showing graphs of $\exv{x}=\mathrm{tr}(\rho_{1}x)$ as a function
of $g_{0}$ for $N=5$. For $\alpha=0$, and of course for $g_{0}=0$,
no modulation exists and, by symmetry, $\exv{x}=0$. Notably, the
same goes for $g_{0}\to\infty$, when the correlation diagonal is
depleted as delineated above, even though the displacement will vanish
only very slowly. There is a trade-off in between for which $\exv{x}$
becomes extremal. The value where this occurs, $g_{0}^{\star}(\alpha)$,
depends only weakly on the relative modulation $\alpha$---despite
the fact that the maximum ground-state displacement $-\exv{x}^{\star}$
will of course increase monotonically with $\alpha$. %

\subsection{Central barrier $h>0$\label{sub:mDW}}

In the presence of a sufficiently strong barrier, the situation is
a different one. To begin with ($g_{0}=0$), the atoms are not centered
as before but rather coherently distributed over the two wells. Hence,
upon switching on the inhomogeneous interaction, they can immediately
feel the full impact of its modulation on both sides. For finite barrier
strength $h$, they can then re-distribute so as to find a compromise
between minimum repulsion and potential energy. Even though this mechanism
is universal for any particle number, we will lay it out for both
even and odd $N$ so as to keep a link to the reference case $\alpha=0$.

\subsubsection{Even $N$}

\begin{figure}
\subfigure[]{\includegraphics[%
  width=6cm,
  keepaspectratio]{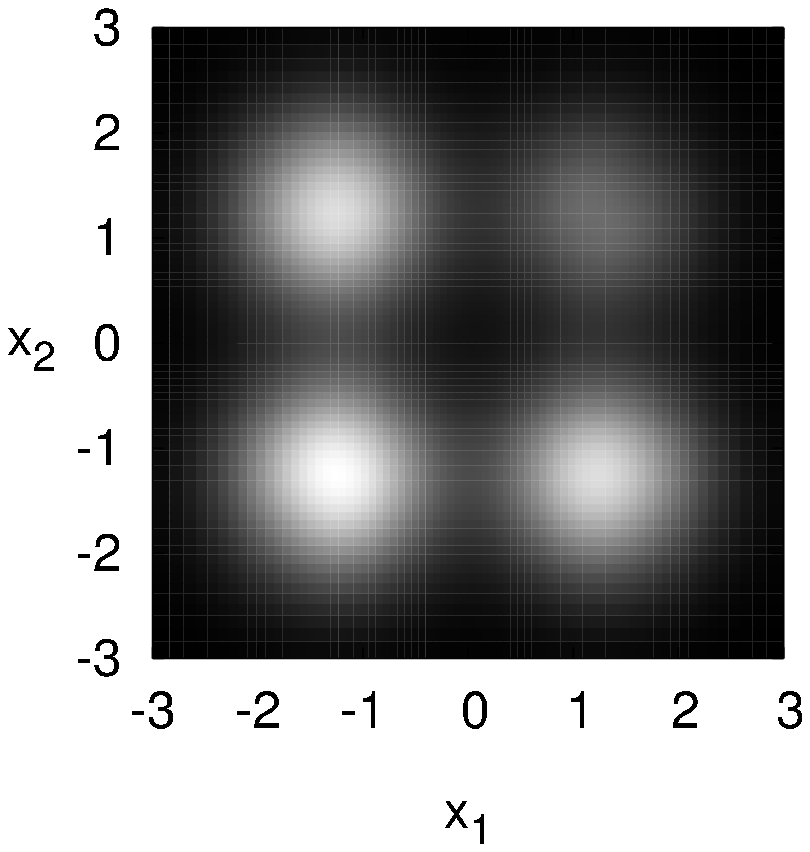}}\subfigure[]{\includegraphics[%
  width=6cm,
  keepaspectratio]{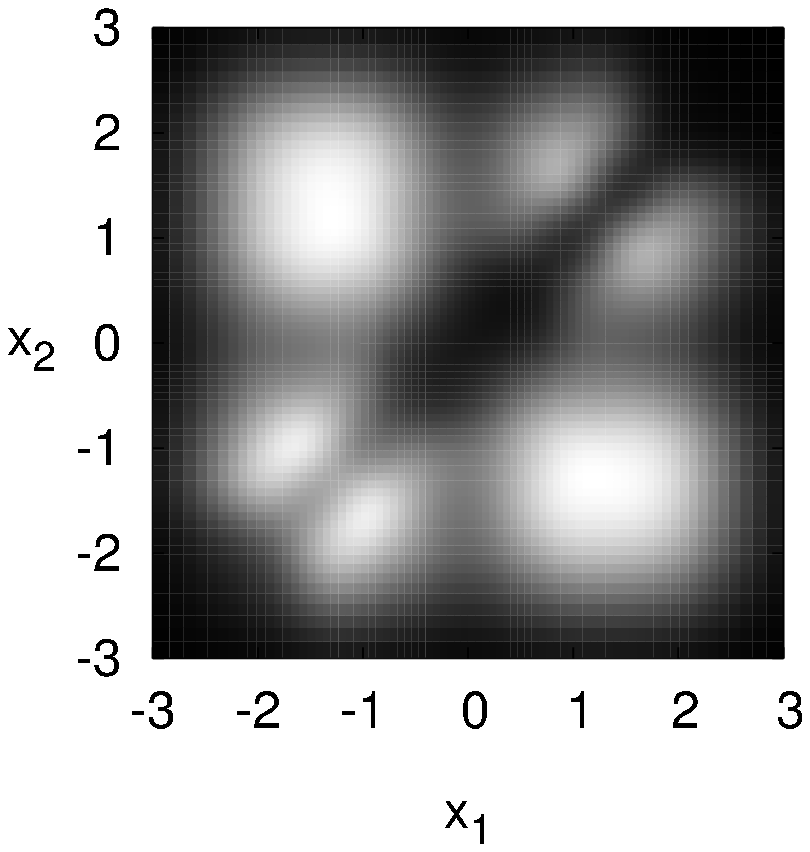}}

\caption{2-particle density for $N=4$ bosons in a double-well trap ($h=5$)
and with inhomogeneous interactions ($\alpha=.5$). (a) Already for
$g=0.2$, the probability of finding any two atoms in the left well
is significantly enhanced. (b) At the onset of fragmentation ($g=15$),
the diagonal $\{ x_{1}=x_{2}\}$ is being depleted. \label{cap:density-mDW}}
\end{figure}

The above process is illustrated in Fig.~\ref{cap:density-mDW},
which evidences an immediate shift from the right well to the left
one, where the repulsion is weaker. This still corresponds to the
Gross-Pitaevskii regime of a single dominant orbital: there is no
correlation hole; in fact the probability density of finding both
particles in the left well, $\rho_{2}(-x_{0},-x_{0})$, may even be
larger than that for separation, $\rho_{2}(\pm x_{0},\mp x_{0})$.
As the interaction passes a critical strength, fragmentation sets
in, somewhat more pronounced on the right-hand side (Fig.~\ref{cap:density-mDW}b).
Note how the diagonal $\{ x_{1}=x_{2}\}$ is being emptied, signifying
the incipient destruction of the imbalance.

This reflects in the 1-particle density displayed in Fig.~\ref{cap:density1-mDW}.
The density is almost {}`instantaneously' shuffled from the right
to the left. In the curve for $g_{0}=4.7$, it becomes apparent that
the fragmentation essentially kicks in separately for both wells,
where only the right well exhibits the typical repulsion-induced split-up.
As asserted already for the harmonic reference case ($h=0$), the
modulation becomes marginal \emph{relative} to the overall fermionization
process. This may also be discerned here: for an even number of bosons,
the asymmetric effect fades, and as in Sec.~\ref{sub:DW}, the fragmentation
is assisted by the central barrier insofar as it now enters separately
in the two wells.

\subsubsection{Odd $N$}

The behavior here is wholly analogous to that evidenced in Fig.~\ref{cap:density-mDW}.
On the correlation diagonal, the density again experiences an asymmetry
even for tiny $g>0$, where the mean-field behavior is still virulent.
For a strong enough modulation, say $\alpha=.5$, coincidence of two
atoms in the left well ($x_{1/2}=-x_{0}$) is even enhanced with respect
to separation ($x_{1}=-x_{2}$), while coincidence in the right well
is extinguished very quickly. At a certain point, fragmentation sets
in, which eventually evolves into fermionization.

\begin{figure}
\subfigure[Four atoms]{\includegraphics[%
  width=8cm,
  keepaspectratio]{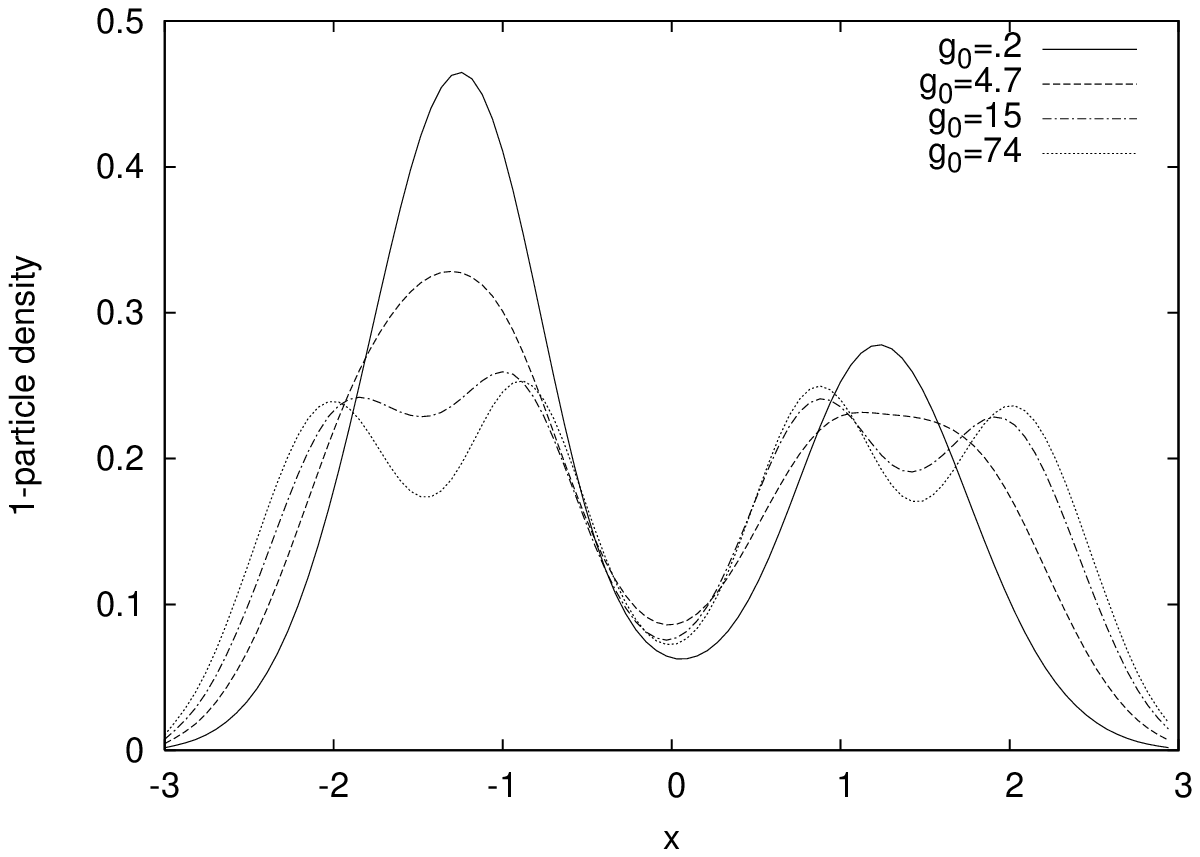}}\subfigure[Five atoms]{\includegraphics[%
  width=8cm,
  keepaspectratio]{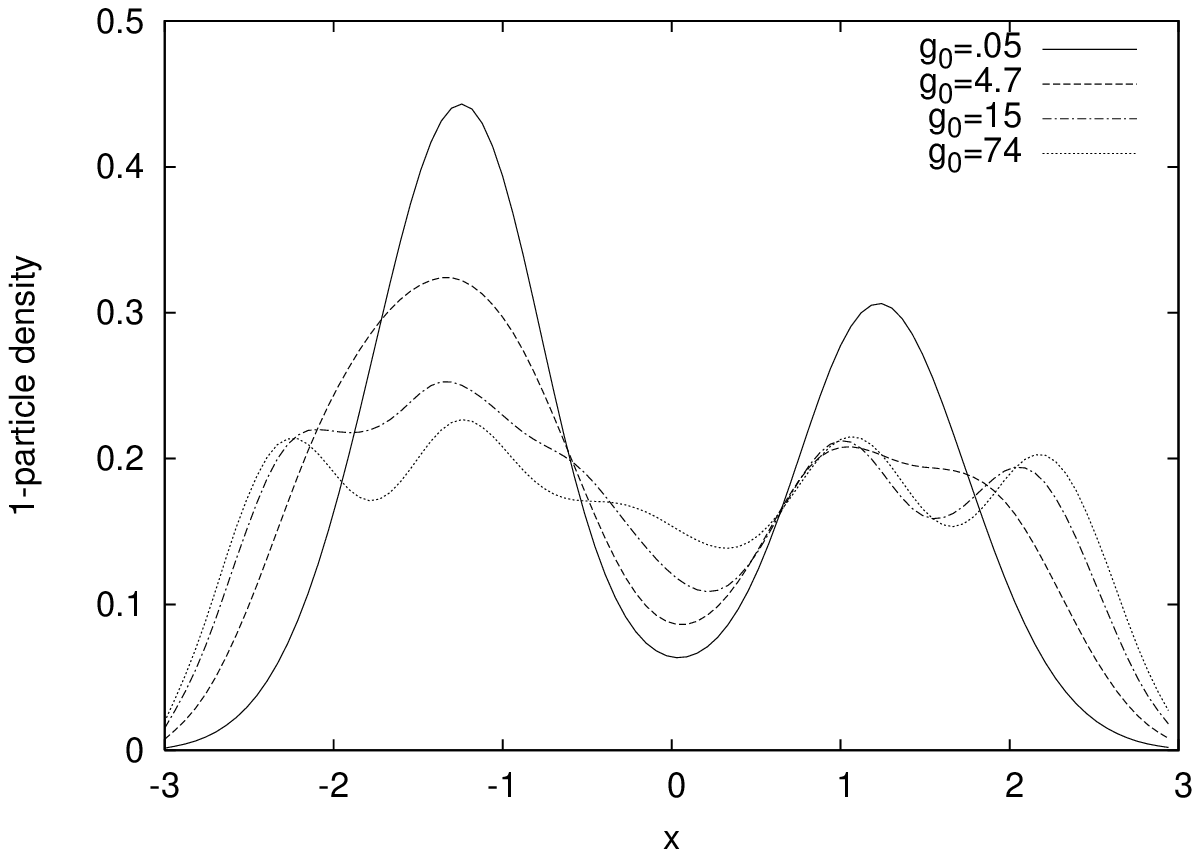}}

\caption{1-particle density for a double well ($h=5$) and modulated coupling
strength ($\alpha=.5$) : Even vs. odd number of atoms. In both cases,
fragmentation essentially sets in separately in each well (e.g., see
$g_{0}=4.7$). However, for $N=4$ (a), the fermionization process
is again supported by the central barrier, while (b) suggests that
for $N=5$ the extra particle can now be temporarily accommodated
by the left well. \label{cap:density1-mDW} }
\end{figure}

It should be noted that the characteristic influence of the even atom
number is on the transition to fermionization. Recall that this was
hampered for \emph{homogeneous} interactions owing to symmetry, which
imposed that one particle had to reside near the center, $x=0$. This
no longer holds here, and in fact, Fig.~\ref{cap:density1-mDW}(b)
unveils that the spare particle is practically accommodated in the
left well.%

The nature of the ground-state displacement is again summarized in
the graph of $-\exv{x}$ (Fig.~\ref{cap:mHO-q}). While the harmonic
system turned out to be rather irresponsive to $g_{0}$, the displacement
now exhibits a dramatic increase with raising $g_{0}$, as laid out
above. It finds a maximum, which corresponds to the trade-off between
localizing in the left well and maximum spreading. As before, the
modulation $\alpha$ does not so much alter the critical $g_{0}^{\star}(\alpha)$,
but of course makes for a stronger maximum displacement $\exv{-x}^{\star}$.
The displacement decreases again slowly beyond that point. A notable
side effect is that the displacement in the presence of a central
barrier may in fact drop \emph{below} the one without it, although
of course this can only happen if the modulation $\alpha$ was smaller
to begin with. That is simply because the double well, favoring the
delocalization of the atoms, not only supports the modulation's effect,
but also accelerates fragmentation and hence---eventually---destruction
of the asymmetry.

\section{Conclusion and outlook}

We have studied the numerically exact ground state of $N$ one-dimensional
bosons in double-well traps, where the atoms interact repulsively
via a short-ranged potential whose \emph{strength} as well as its
\emph{spatial modulation} could be tuned. Our approach relied on the
Multi-Configuration Time-Dependent Hartree method, a wave-packet dynamics
tool known for its efficiency in higher dimensions. This allowed us
to study the interplay between different heights of the barrier separating
both wells, on the one hand, and different interaction strengths as
well as spatial modulations in all relevant regimes in a numerically
exact fashion.

For standard, homogeneous interactions, we have witnessed the transition
from the weak-coupling mean-field regime to fragmentation and finally
to a fermionized ground state for very large repulsion. Absent any
barrier, this process requires sufficient interaction energies so
as to compensate the one-particle (kinetic and potential) energy added
by a fragmentation; also see \cite{alon05}. Its signature is a broadening
and eventually the appearance of $N$ humps in the density profile,
plus a {}`correlation hole' in the two-particle density. As demonstrated,
this also reflects in the relative occupation of the dominant natural
orbital, $n_{0}$, which reduces from unity to order of $1/\sqrt{N}$
as the interaction is increased. As we turn the harmonic trap into
a double well, then well below the barrier energy the fragmentation
essentially takes place separately in each well, whereas way above
the barrier, the situation resembles that in the harmonic trap. In
particular, we find that for even $N$, fermionization is assisted,
while it is impeded for odd $N$, when by symmetry one particle should
be distributed over the barrier region.

We have also tackled the question of inhomogeneous effective interactions,
insomuch as the coupling strength is assumed to be larger on one side
of the trap. We have found the ground state to be displaced toward
the side where repulsive forces are weaker. This displacement can
be enhanced by stronger modulations, whereas the optimal interaction
strength needed to achieve it can be lowered primarily by higher barriers.
It falls off for stronger interactions as the fermionization destructs
any imbalance effects.\\

This work can be seen as a model study of the effects of inhomogeneous
interactions, in particular a setup with an imbalance inducing a parity
violation. It can in principle be extended to higher particle numbers,
although this was not done here with an eye toward time and computational
effort. On the other hand, there are also plenty of promising other
configurations that come into question, such as enhancing the transmission
through a barrier by modulating the interaction strength accordingly
or using more than two wells so as to separate single atoms from the
system. Of course, to address these problems, eventually a time-dependent
simulation is mandatory to gain insight into realistic situations,
while the extension to more dimensions may become inevitable. 

\begin{acknowledgments}
Financial support from the Landesstiftung Baden-Württemberg in the
framework of the project {}`Mesoscopics and atom optics of small
ensembles of ultracold atoms' is gratefully acknowledged by PS and
SZ. These two authors also appreciate A. Streltsov's helpful comments
and thank O. Alon for illuminating discussions. 
\end{acknowledgments}
\bibliographystyle{prsty}
\bibliography{/home/sascha/paper/pra/DW/phd,/home/sascha/bib/mctdh}

\begin{thebibliography}{10}

\bibitem{pitaevskii}
L. Pitaevskii and S. Stringari, {\em Bose-Einstein Condensation} (Oxford
  University Press, Oxford, 2003).

\bibitem{dalfovo99}
F. Dalfovo, S. Giorgini, L. Pitaevskii, and S. Stringari, Rev. Mod. Phys. {\bf
  71},  463  (1999).

\bibitem{pethick}
C.~J. Pethick and H. Smith, {\em Bose-Einstein condensation in dilute gases}
  (Cambridge University Press, Cambridge, 2001).

\bibitem{leggett01}
A.~J. Leggett, Rev. Mod. Phys. {\bf 73},  307  (2001).

\bibitem{Olshanii1998a}
M. Olshanii, Phys. Rev. Lett. {\bf 81},  938  (1998).

\bibitem{girardeau60}
M. Girardeau, J. Math. Phys. {\bf 1},  516  (1960).

\bibitem{kinoshita04}
T. Kinoshita, T. Wenger, and D.~S. Weiss, Science {\bf 305},  1125  (2004).

\bibitem{paredes04}
B. Paredes {\it et~al.}, Nature {\bf 429},  277  (2004).

\bibitem{cederbaum03}
L.~S. Cederbaum and A.~I. Streltsov, Phys. Lett. A {\bf 318},  564  (2003).

\bibitem{alon05}
O.~E. Alon and L.~S. Cederbaum, Phys. Rev. Lett. {\bf 95},  140402  (2005).

\bibitem{Busch98}
{Th. Busch, B. G. Englert, K. Rzazewski, and M. Wilkens}, Found. Phys. {\bf
  28},  549  (1998).

\bibitem{cirone01}
M.~A. Cirone, K. G\'{o}ral, K. Rzazewski, and M. Wilkens, J. Phys. B {\bf 34},
  4571  (2001).

\bibitem{hao06}
Y. Hao, Y. Zhang, J.~Q. Liang, and S. Chen, cond-mat/0602483  (2006).

\bibitem{sakmann05}
K. Sakmann, A.~I. Streltsov, O.~E. Alon, and L.~S. Cederbaum, Phys. Rev. A {\bf
  72},  033613  (2005).

\bibitem{idziaszek06}
Z. Idziaszek and T. Calarco, quant-ph/0604205  (2006).

\bibitem{Blume2002a}
D. Blume and C.~H. Greene, Phys. Rev. A {\bf 65},  043613  (2002).

\bibitem{bolda03}
E.~L. Bolda, E. Tiesinga, and P.~S. Julienne, Phys. Rev. A {\bf 68},  032702
  (2003).

\bibitem{tiesinga00}
F.~M. E.~Tiesinga, C.J.~Williams and P. Julienne, Phys. Rev. A {\bf 61},
  063416  (2000).

\bibitem{idziaszek:050701}
Z. Idziaszek and T. Calarco, Phys. Rev. A {\bf 71},  050701  (2005).

\bibitem{masiello05}
D. Masiello, S.~B. McKagan, and W.~P. Reinhardt, Phys. Rev. A  063624  (2005).

\bibitem{streltsov06}
A.~I. Streltsov, O.~E. Alon, and L.~S. Cederbaum, cond-mat/0603212  (2006).

\bibitem{deuretzbacher06}
F. Deuretzbacher, K. Bongs, K. Sengstock, and D. Pfannkuche, cond-mat/0604673
  (2006).

\bibitem{sundholm04}
D. Sundholm and T. V{\"a}nsk{\"a}, J. Phys. B {\bf 37},  2933  (2004).

\bibitem{haugset97}
T. Haugset and H. Haugerud, Phys. Rev. A {\bf 57},  3809  (1998).

\bibitem{klaiman06}
S. Klaiman, N. Moiseyev, and L.~S. Cederbaum, Phys. Rev. A {\bf 73},  013622
  (2006).

\bibitem{mey03:251}
H.-D. Meyer and G.~A. Worth, Theor.\ Chem.\ Acc. {\bf 109},  251  (2003).

\bibitem{mey98:3011}
H.-D. Meyer,  in {\em {T}he {E}ncyclopedia of {C}omputational {C}hemistry},
  edited by P. v.~R.~Schleyer {\it et~al.} (John Wiley and Sons, Chichester,
  1998), Vol.~5, pp.\ 3011--3018.

\bibitem{bec00:1}
M.~H. Beck, A. J{\"a}ckle, G.~A. Worth, and H.-D. Meyer, Phys.\ Rep. {\bf 324},
   1  (2000).

\bibitem{kuhr01}
S. Kuhr {\it et~al.}, Science {\bf 293},  278  (2001).

\bibitem{mohring05}
B. Mohring {\it et~al.}, Phys. Rev. A {\bf 71},  053601  (2005).

\bibitem{huang57}
K. {Huang} and C.~N. {Yang}, Phys. Rev. {\bf 105},  767  (1957).

\bibitem{mey90:73}
H.-D. Meyer, U. Manthe, and L.~S. Cederbaum, Chem.\ Phys.\ Lett. {\bf 165},  73
   (1990).

\bibitem{Penrose56}
O. Penrose and L. Onsager, Phys. Rev. {\bf 104},  576  (1956).

\bibitem{mctdh:package}
G.~A. Worth, M.~H. Beck, A. J{\"a}ckle, and H.-D. Meyer, The {MCTDH} {P}ackage,
  {V}ersion 8.2, (2000). H.-D. Meyer, {V}ersion 8.3 (2002). {S}ee
  http://www.pci.uni-heidelberg.de/tc/usr/mctdh/.

\bibitem{kos86:223}
R. Kosloff and H. Tal-Ezer, Chem.\ Phys.\ Lett. {\bf 127},  223  (1986).

\bibitem{Busch03}
T. Busch and G. Huyet, J. Phys. B {\bf 36},  2553  (2003).

\bibitem{girardeau01}
M. Girardeau, E.~M. Wright, and J.~M. Triscari, Phys. Rev. A {\bf 63},  033601
  (2001).

\bibitem{lieb03}
E.~H. Lieb, R. Seiringer, and J. Yngvason, Phys. Rev. Lett. {\bf 91},  150401
  (2003).

\bibitem{theocharis05}
G. Theocharis, P. Schmelcher, P.~G. Kevrekidis, and D.~J. Frantzeskakis,
  cond-mat/0505127  (2005).

\end{thebibliography}

\end{document}